\documentclass[aps,prb,twocolumn,showpacs,superscriptaddress,floatfix]{revtex4-1}
\usepackage{bm}
\usepackage{color}
\usepackage{amssymb}
\usepackage{graphicx}
\usepackage{amsmath}
\usepackage{dcolumn}
\usepackage{bm}
\usepackage{soul}
\usepackage[german,english]{babel}

\begin{document}
\title{Transport and Optical Properties of Single- and Bilayer Black Phosphorus with Defects}
\author{Shengjun Yuan}
\email{s.yuan@science.ru.nl}
\affiliation{Radboud University, Institute for Molecules and Materials,
Heijendaalseweg 135, 6525 AJ Nijmegen, The Netherlands}
\author{A. N. Rudenko}
\affiliation{Radboud University, Institute for Molecules and Materials,
Heijendaalseweg 135, 6525 AJ Nijmegen, The Netherlands}
\author{M. I. Katsnelson}
\affiliation{Radboud University, Institute for Molecules and Materials,
Heijendaalseweg 135, 6525 AJ Nijmegen, The Netherlands}
\pacs{73.22.-f, 71.23.-k, 78.20.Bh, 61.43.Bn}
\date{\today }

\begin{abstract}
We study the electronic and optical properties of single- and bilayer black phosphorus
with short- and long-range defects by using the tight-binding propagation method.
Both types of defect states are localized and induce a strong scattering 
of conduction states reducing significantly the charge carrier mobility. 
In contrast to properties of pristine samples, the anisotropy of defect-induced optical 
excitations is suppressed due to the isotropic nature of the defects. 
We also investigate the Landau level spectrum and magneto-optical conductivity,
and find that the discrete Landau levels are sublinearly dependent 
on the magnetic field and energy level index, even at low defect concentrations.
\end{abstract}

\maketitle

\section{Introduction}

Black phosphorus (BP) is a layer material in which the atomic layers are
coupled by weak van der Waals interactions. Few-layer BP is a new kind of
two-dimensional (2D) material that can be obtained by mechanical exfoliation
method from BP films\cite%
{LLi2014,HLiu2014,XiaF2014,Koenig2014,Castellanos2014,LiL2014}, a common fabricating
method of producing 2D materials. BP is a semiconductor with layer-dependent
direct band gap, which is crucial for a
number of applications such as field-effect transistors\cite{Qiao2014,Rudenko2014,Tran2014,GuanJ2014,PengXH2014}. The anisotropic
optical response of BP\cite{XiaF2014,Qiao2014,Tran2014,Cakir2014,Low2014,Low2014b,LiPK2014}, which is
not typical for other known 2D materials, makes it an ideal material for
photon polarizer. On the other hand, unlike graphene and transitionmetal
dichalcogenides, which are chemically stable under ambient conditions, BP
samples are shown to be very sensitive to the environment\cite%
{Koenig2014,Castellanos2014,Favron2014,Wood2014,Island2015}. This is due to the high reactivity
of BP with respect to air and might limit their application in real devices.
In this regard, the role of defects and impurities in BP represents an
important issue with theoretical and practical relevance.

\begin{figure}[t]
\begin{center}
\mbox{
\includegraphics[width=8cm]{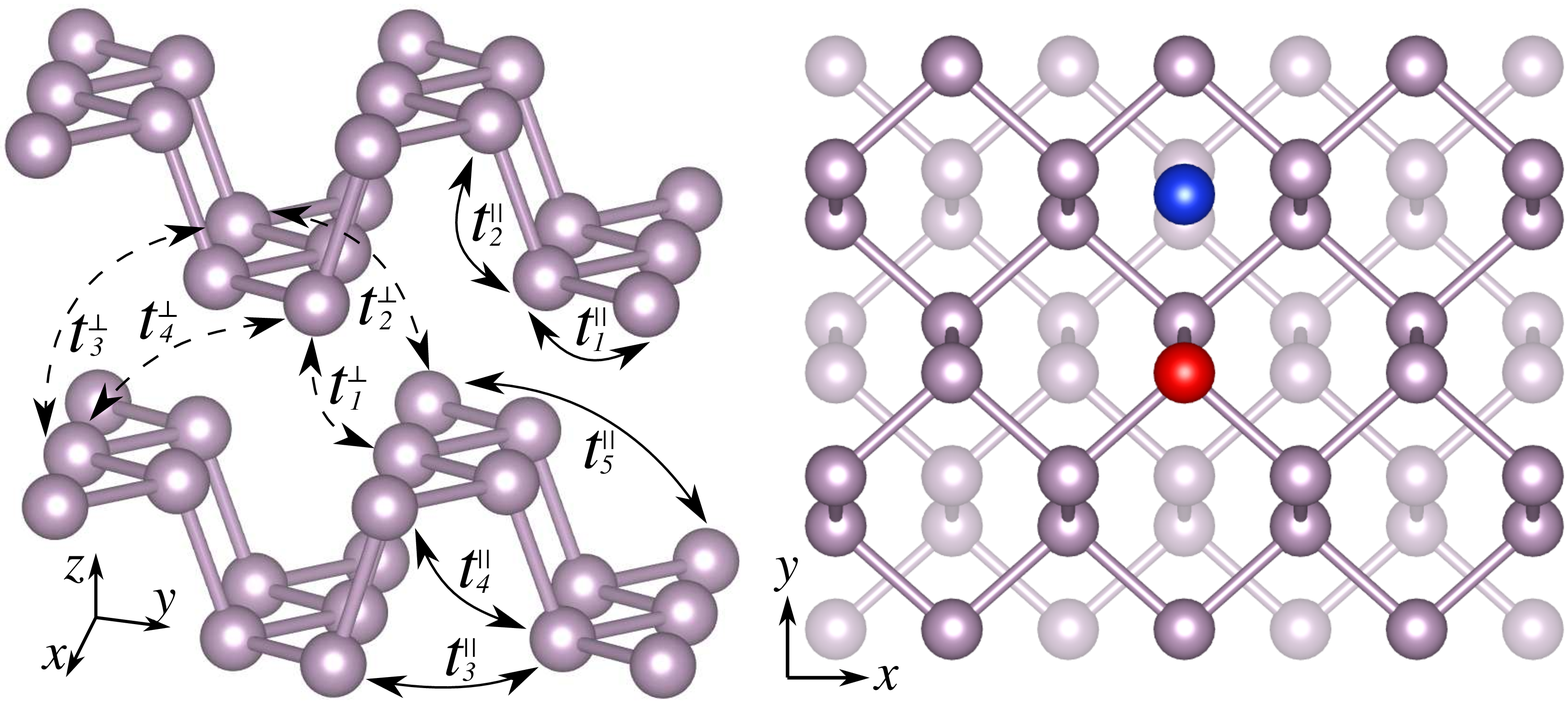}
} \vspace{0.5em}
\mbox{
\includegraphics[width=4.25cm]{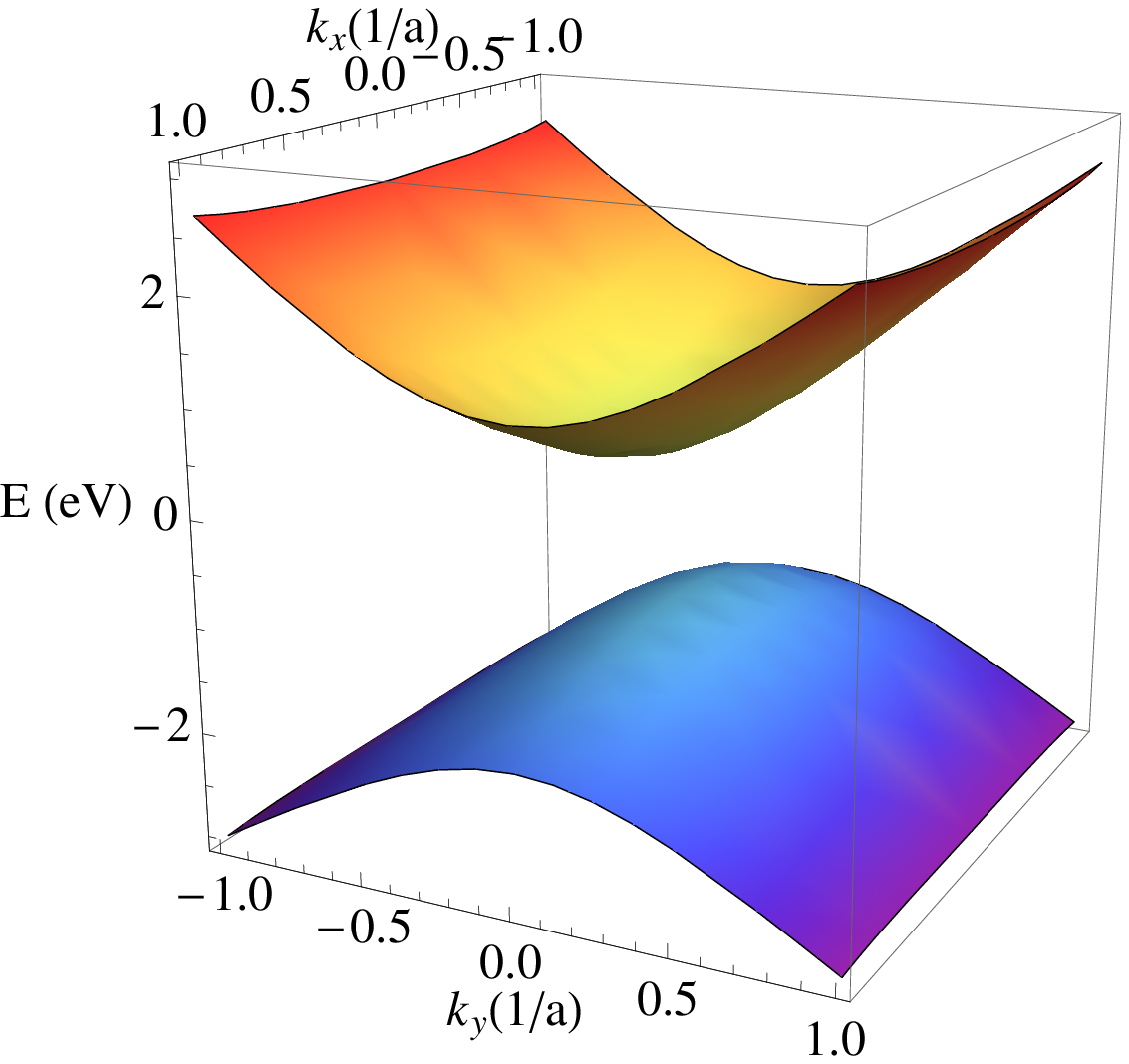}
\includegraphics[width=4.25cm]{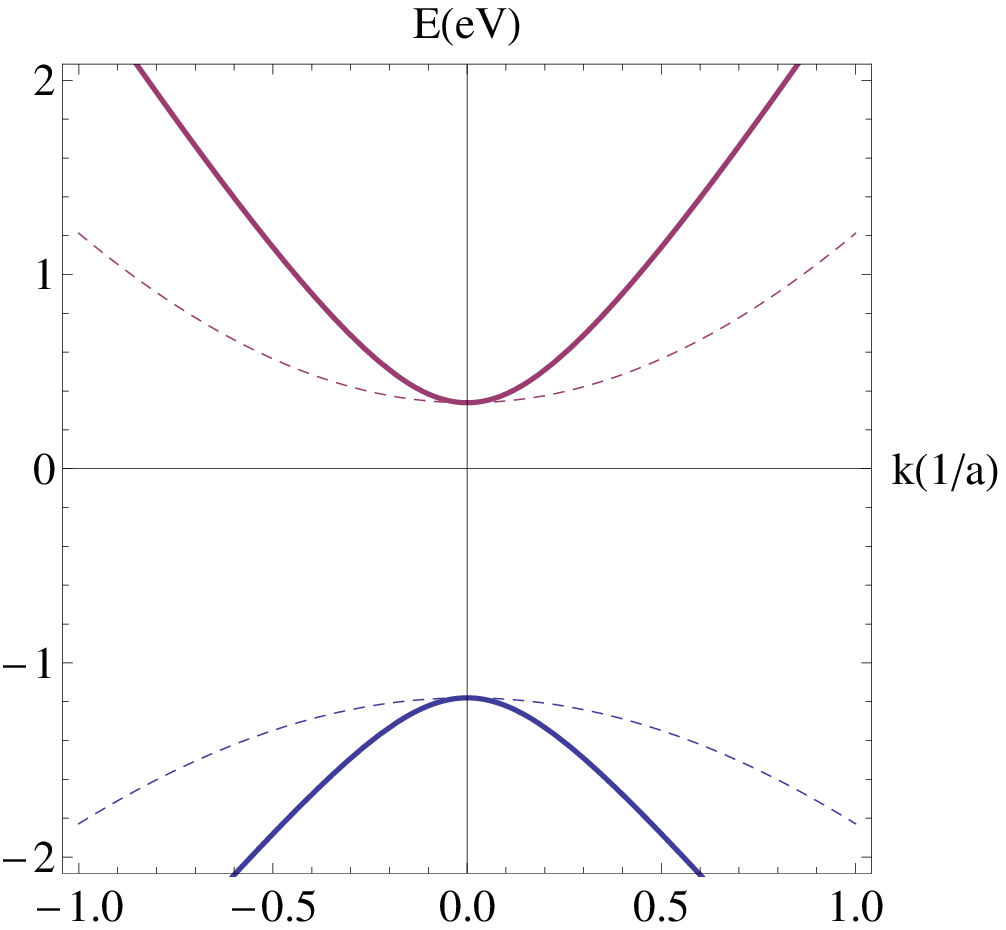}
}
\end{center}
\caption{Top-left: Schematic representation of the atomic structure and
hopping parameters of the TB model of BP. Top-right: Top view of the atomic
structure, the red dot represents a point defect, and
the blue dot a long-range defect at the center of the projected
honeycomb lattice on the surface. Bottom-left: 3D contour plot of the lowest
valence and conduction bands. Bottom-right: Lowest valence and conduction
bands along armchair (solid lines) and zigzag (dashed lines) directions.}
\label{Fig:Hoppings_Band}
\end{figure}

In 2D materials, the scattering induced by short-range point defects (like adsorbates in
graphene and sulphur vacancies in MoS$_{2}$) are shown to be one of the main
mechanism dominating the charge mobility\cite%
{Neto2009,Sarma2011,Katsnelsonbook,WK10,YRK10,YRK10b,QiuH2013,Yuan2014}. The
point defects are the so called resonant scatterers as they can provide
resonances (quasilocalized states) near the neutrality point (in graphene 
\cite{WK10,YRK10}) or within the band gap (in the semiconducting
transitionmetal dichalcogenides such as MoS$_{2}$ and WS$_{2}$ \cite%
{Yuan2014,QiuH2013}). The emergence of the midgap states due to  
point defects in BP has been observed in several first-principles calculations, such as single
vacancy\cite{Liu2014,HuW2014}, substitutional p-dopants (Te, C)\cite%
{Liu2014}, oxygen bridge-type defects\cite{Ziletti2015},
absorption of organic molecules\cite{Zhang2014} or adatoms (Si,
Ge, Au, Ti, V)\cite{Kulish2015}. But it remains unclear what is the influence of the resonant point defects
to the transport and optical properties of BP, as the first-principles calculations are limited by
the sample size that is computationally too expensive to consider a large sample with many defects.
% There are also other types of point defects 
% without the formation of midgap states\cite{Liu2014,HuW2014,Ziletti2015,Zhang2014,Kulish2015,WangV2014},
% which are beyond the scope of our current study.
Another typical disorder in 2D materials is the long-range
electron-hole puddles\cite{Neto2009,Sarma2011,Katsnelsonbook}, which are
inhomogeneities of carrier-density and have been observed experimentally\cite%
{Martin2007,Neto2009}. The origin of electron-hole puddles could be charged
impurities and defects located on the substrate \cite%
{Hwang2007,ZhangYB2009,Rudenko2011} or surface corrugations such as ripples
and wrinkles\cite{Gibertini2010,Gibertini2012}. Unlike the point defects,
the electron-hole puddles do not introduce strong resonances in the spectrum
and therefore are referred to as typical non-resonant defects. In this
letter, the study of disordered samples are performed by using the
tight-binding propagation method (TBPM) \cite{YRK10,WK10,YRRK11,Yuan2012},
which is extremely efficient in the large-scale calculation of systems with
more than millions of atoms.

% The TB Hamiltonian for pristine BP is based on the $GW$ approximation\cite{Rudenko2014},
% providing accurate description of its conduction and valence bands in the ranges of $\sim $0.3 eV beyond
% the gap (for details of the TB Hamiltonian we refer to Ref. \cite{Rudenko2014}).
\section{Tight-binding Model}
The TB Hamiltonian for pristine BP is based on the $GW$ approximation,
and it follows\cite{Rudenko2014}:%
\begin{equation}
H=\sum_{i}\varepsilon _{i}n_{i}+\sum_{i\neq j}t_{ij}c_{i}^{\dag
}c_{j}+\sum_{i\neq j}t_{ij}^{\perp }c_{i}^{\dag }c_{j},  \label{tb_hamilt}
\end{equation}%
where the summation runs over the lattice sites of single- or bilayer BP, $%
\varepsilon _{i}$ is the energy of the electron at site $i$, $t_{ij}^{(\perp
)}$ is the intralayer (interlayer) hopping parameter between the $i$th and $%
j $th sites, and $c_{i}^{\dag }$ ($c_{j}$) is the creation (annihilation)
operator of electrons at site $i$ ($j$). The parameters $t_{ij}$, $%
t_{ij}^{\perp }$, and $\epsilon _{i}$ were obtained on the basis of accurate 
\textit{ab initio} calculations within the $G_{0}W_{0}$ approximation by
mapping the entire manifold of $sp$ states onto the minimal set (one site
per P atom) of relevant states near the band gap.
Specifically, we use five intralayer ($t_{1}=-1.220$ eV, $t_{2}=3.665$ eV, $%
t_{3}=-0.205$ eV, $t_{4}=-0.105$ eV, $t_{5}=-0.055$ eV) and four interlayer
hoppings ($t_{1}^{\perp }=0.295$ eV, $t_{2}^{\perp }=0.273$ eV, $%
t_{3}^{\perp }=-1.151$ eV, $t_{4}^{\perp }=-0.091$ eV), which schematically
% shown in Fig. 1 of Supplementary Materials, and an energy splitting of $\Delta
shown in Fig.~\ref{Fig:Hoppings_Band}, and an energy splitting of $\Delta
\epsilon =1.0$eV between the nonequivalent electrons in bilayer BP.\cite{Rudenko2014}. The
resulting TB model accurately describes the quasiparticle electron and hole
bands of single layer and bilayer BP in the ranges of $\sim $0.3 eV beyond
the gap\cite{Rudenko2014}.

\begin{figure}[t]
\begin{center}
\mbox{
\includegraphics[width=4cm]{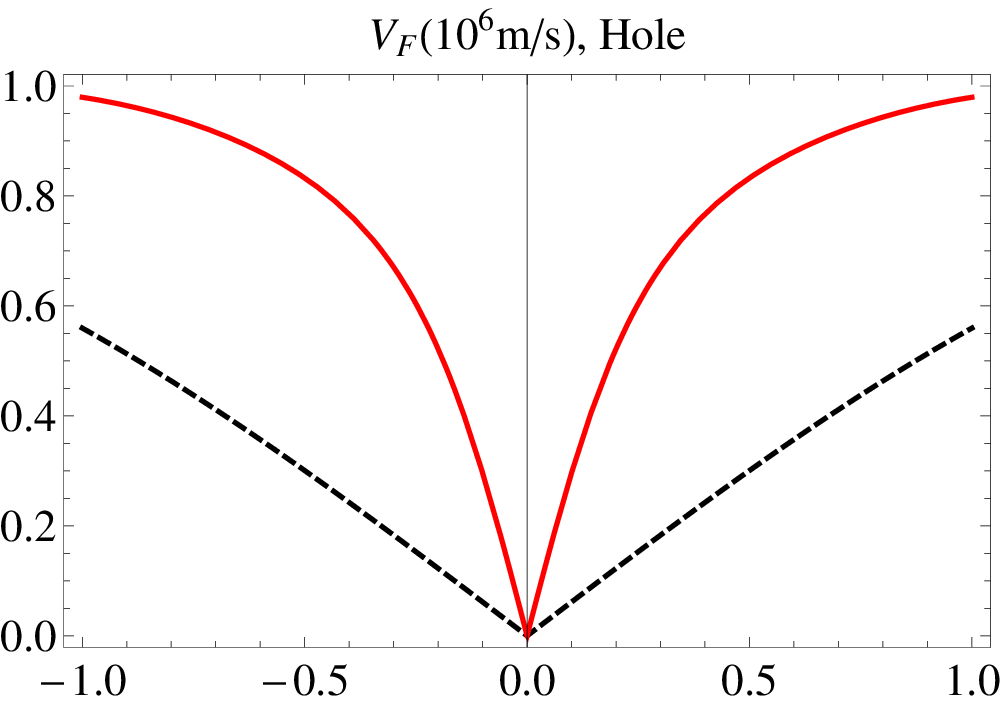}
\includegraphics[width=4cm]{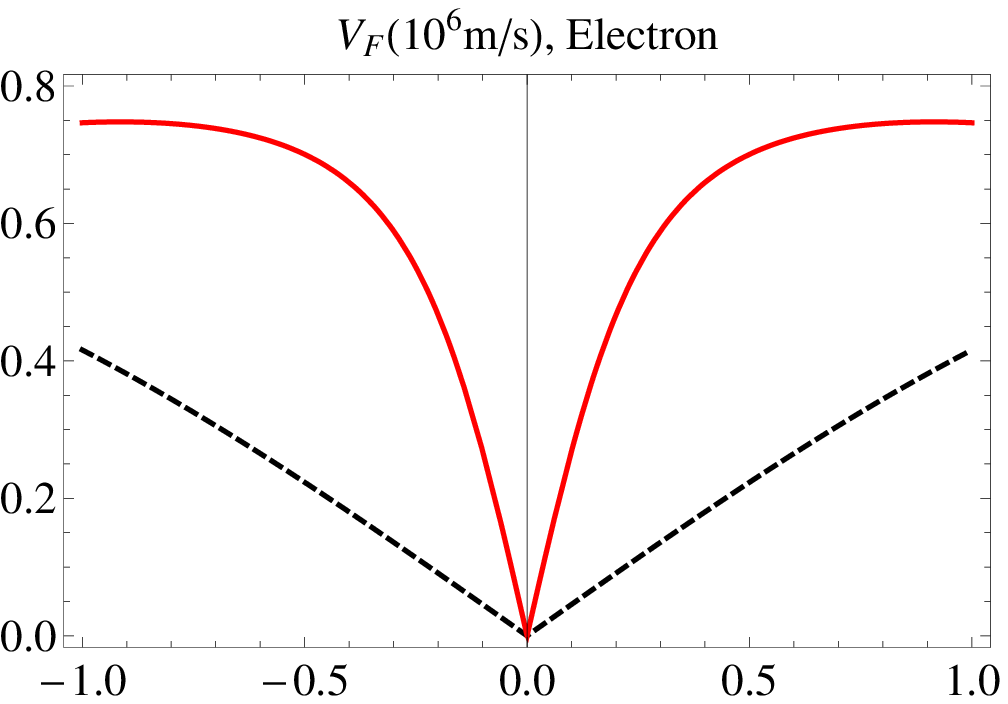}
} 
\mbox{
\includegraphics[width=4.25cm]{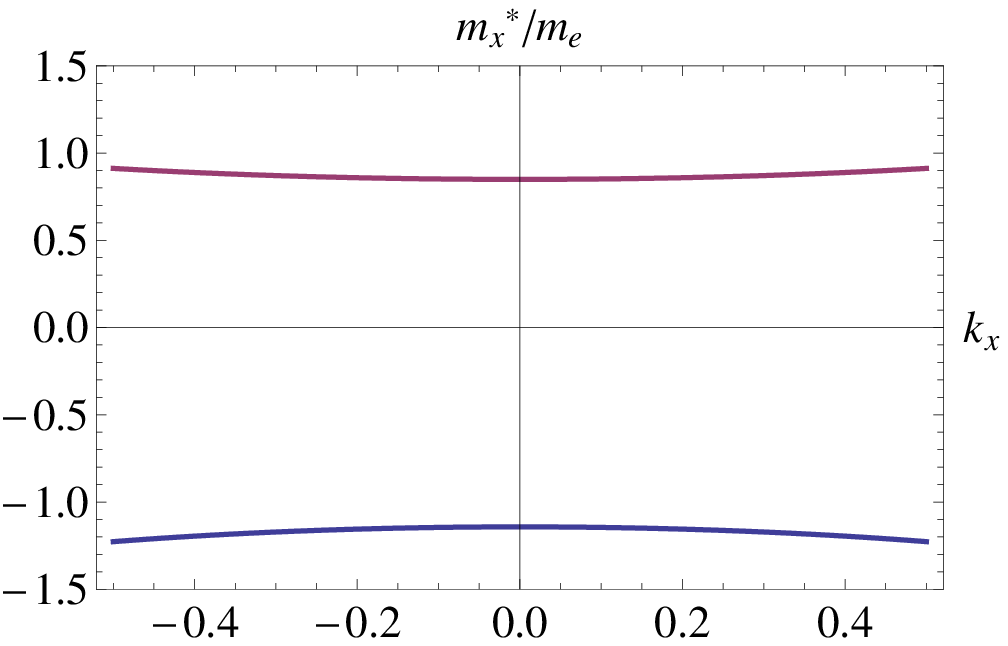}
\includegraphics[width=4.25cm]{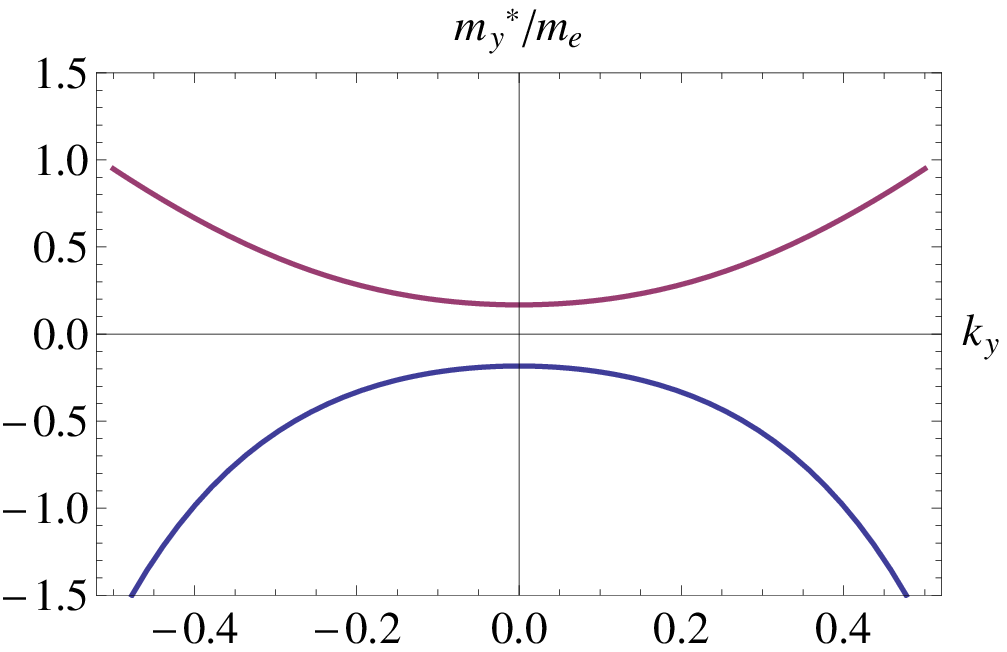}
}
\end{center}
\caption{Top: Electron and hole velocities in single-layer BP as a function
of wave vector along armchair (red solid line) and zigzag (black dashed
line) directions. Bottom: Effective mass along armchair (Y) and zigzag (X)
directions. The wave vector $k$ is in units of $1/a$, where $a\approx 2.216$\AA\ is the atomic
distance between two nearest neighbors.}
\label{Fig:Velocity_Mass}
\end{figure}

The energy dispersions $E(k_{x},k_{y})$ can be obtained analytically by diagonalizing the TB
Hamiltonian.
(see the band structure plotted in Fig.~\ref{Fig:Hoppings_Band}
and detailed calculations in the Appendix). 
% (see the band structure plotted in Supplementary Materials). 
% The anisotropy of BP is identified directly from the anisotropic Fermi velocities 
% $v_{\alpha }=\frac{1}{\hbar }\frac{\partial E}{\partial k_{\alpha }}$ and
% effective masses $m_{\alpha }=\hbar ^{2}/\frac{\partial ^{2}E}{\partial
% k_{\alpha }^{2}}$ shown in Fig.~\ref{Fig:Velocity_Mass}. 
The anisotropy can be further identified directly from the anisotropic Fermi velocities and
effective masses shown in Fig.~\ref{Fig:Velocity_Mass}. 
The Fermi velocity $%
v_{\alpha }=\frac{1}{\hbar }\frac{\partial E}{\partial k_{\alpha }}$ and
effective mass $m_{\alpha }=\hbar ^{2}/\frac{\partial ^{2}E}{\partial
k_{\alpha }^{2}}$ are calculated from the energy dispersion relations (see details in the Appendix). 
The velocity of an electron (hole) along the armchair direction is much larger
than the value along the zigzag direction, with both linear $k$-dependent
velocities around the $\Gamma $ point ($k=0$). This is different from the
Dirac fermion in graphene which travels with constant velocity $v_{F}\simeq
c/300$, where $c$ is the speed of light in vacuum. The effective masses at
the $\Gamma $ point along the armchair direction are $m_{y}^{v}=0.184m_{e}$
for hole, and $m_{y}^{c}=0.167m_{e}$ for electron. Here, $m_{e}$ is the
free electron mass. The effective masses along the zigzag direction are much
heavier: $m_{x}^{v}=1.143m_{e}$ for hole and $m_{x}^{c}=0.849m_{e}$ for
electron. 
% Our analysis of pristine BP shows clearly that anisotropy is an intrinsic
% property of BP due to its unique band structure. 

\section{Tight-binding Propagation Method}

The electronic and optical properties of single- and bilayer BP are
calculated by using the TBPM,\cite{HR00,WK10,YRK10,Yuan2012} in which the
initial state $\left\vert \varphi \right\rangle $ is chosen as a random
superposition of all sites over the whole space which covers all the energy
eigenstates \cite{HR00,YRK10} $\left\vert \varphi \right\rangle
=\sum_{i}a_{i}\left\vert i\right\rangle ,$where $a_{i}$ are random complex
numbers normalized as $\sum_{i}\left\vert a_{i}\right\vert ^{2}=1$, and $%
\left\vert i\right\rangle $ represents basis state at site $i$. The density
of states can be obtained by the Fourier transformation of the overlap
between the time-evolved state $\left\vert \varphi (t)\right\rangle \equiv
e^{-i\mathcal{H}t}\left\vert \varphi \right\rangle $ and the initial state $%
\left\vert \varphi \right\rangle $ as \cite{HR00,YRK10} 
\begin{equation}
\rho \left( \varepsilon \right) =\frac{1}{2\pi }\int_{-\infty }^{\infty
}e^{i\varepsilon t}\left\langle \varphi |\varphi (t)\right\rangle dt.
\label{Eq:DOS}
\end{equation}%
Here we use units such that $\hbar =1$. The time-evolution operator $e^{-i%
\mathcal{H}t}$ is calculated numerically by using Chebyshev polynomial
algorithm, extremely efficient for a TB Hamiltonian $\mathcal{H}$ which is a
sparse matrix. Within the TBPM, the optical conductivity (omitting the Drude
contribution at $\omega =0$) is calculated by using the Kubo formula \cite%
{Ishihara1971,YRK10} 
\begin{eqnarray}
\sigma _{\alpha \beta }\left( \omega \right) &=&\lim_{\epsilon \rightarrow
0^{+}}\frac{e^{-\tilde{\beta}\omega }-1}{\omega \Omega }\int_{0}^{\infty
}e^{-\epsilon t}\sin \omega t  \notag  \label{gabw2} \\
&&\times 2~\text{Im}\left\langle \varphi |f\left( \mathcal{H}\right)
J_{\alpha }\left( t\right) \left[ 1-f\left( \mathcal{H}\right) \right]
J_{\beta }|\varphi \right\rangle dt,  \notag \\
&&  \label{Eq:OptCond}
\end{eqnarray}%
where $\tilde{\beta}=1/k_{B}T$ is the inverse temperature, $\Omega $ is the
sample area, $f\left( \mathcal{H}\right) =1/\left[ e^{\tilde{\beta}\left( 
\mathcal{H}-\mu \right) }+1\right] $ is the Fermi-Dirac distribution
operator, and the time-dependent current operator in the $\alpha $ ($=x$ or $%
y$) direction is defined as $J_{\alpha }\left( t\right) =e^{i\mathcal{H}%
t}J_{\alpha }e^{-i\mathcal{H}t}$. 

The optical conductivity at an arbitrary direction follows%
\begin{equation}
\sigma _{\theta }\left( \omega \right) =\sigma _{xx}\left( \omega \right)
\cos ^{2}\theta +\sigma _{yy}\left( \omega \right) \sin ^{2}\theta ,
\label{Eq:OptAngle}
\end{equation}%
where $\theta $ is the angle between the polarized direction and $x$ axis.
Eq. (\ref{Eq:OptAngle}) can be derived from the Kubo formula by using the
relation of the current operator $J_{\theta }=J_{x}\cos \theta +J_{y}\sin
\theta $.

The reflection and transmission of a polarized light through BP film can be
solved by using the Maxwell equations with a conducting layer. For the case
of normal incidence, the reflectivity can be expressed as \cite{JacksonBook}%
\begin{equation}
r_{\theta }\left( \omega \right) =-\frac{\varepsilon _{0}c\left( \sqrt{%
\varepsilon _{2}}-\sqrt{\varepsilon _{1}}\right) +\sigma _{\theta }\left(
\omega \right) }{\varepsilon _{0}c\left( \sqrt{\varepsilon _{2}}+\sqrt{%
\varepsilon _{1}}\right) +\sigma _{\theta }\left( \omega \right) },
\end{equation}%
where $\varepsilon _{0}$ is the permittivity of vacuum, $\varepsilon _{1}$
and $\varepsilon _{2}$ the relative permittivity of two media on the two
sides of BP film, and $c$ is the speed of light. The reflection and
transmission probabilities are given by $\mathcal{R}\approx \left\vert
r\right\vert ^{2}$ and $\mathcal{T}\approx \left\vert 1+r\right\vert ^{2}%
\sqrt{\varepsilon _{2}/\varepsilon _{1}}$, and the absorption coefficient
follows $\alpha =1-\mathcal{R-T}$. The absorption coefficient of BP films
can be obtained directly in the optical measurements, such as FTIS \cite%
{XiaF2014,Low2014}. In this work, we show the results with $%
\varepsilon _{1}=\varepsilon _{2}=1$ in order to ignore the influence of the
dielectric substrate.

The dc conductivity at energy $E=\varepsilon $ is calculated by using the
Kubo formula at $\omega \rightarrow 0$ \cite{Ishihara1971,YRK10} 
\begin{eqnarray}
\mathbf{\sigma }_{\alpha \alpha } &=&\underset{\tau \rightarrow \infty }{%
\lim }\mathbf{\sigma }_{\alpha \alpha }\left( \tau \right)  \notag
\label{Eq:DC} \\
&=&\underset{\tau \rightarrow \infty }{\lim }\frac{\rho \left( \varepsilon
\right) }{\Omega }\int_{0}^{\tau }dt~\text{Re}\left[ e^{-i\epsilon
t}\left\langle \varphi \right\vert J_{\alpha }e^{i\mathcal{H}t}J_{\alpha
}\left\vert \varepsilon \right\rangle \right] ,  \notag \\
&&
\end{eqnarray}%
where $\left\vert \varepsilon \right\rangle $ is the \textit{normalized}
quasi-eigenstate \cite{YRK10,YRRK11,Yuan2012}. The semi-classic dc
conductivity $\sigma _{sc}$ without considering the effect of Anderson
localization is defined as the maximum of $\mathbf{\sigma }_{\alpha \alpha
}\left( \tau \right) $ obtained from the integral in Eq. (\ref{Eq:DC}). The
measured field-effect carrier mobility is related to the semi-classic dc
conductivity as $\mu \left( E\right) =\sigma _{sc}\left( E\right)
/en_{e}\left( E\right) $, where the carrier density $n_{e}\left( E\right) $
is obtained from the integral of density of states via $n_{e}\left( E\right)
=\int_{0}^{E}\rho \left( \varepsilon \right) d\varepsilon $.

We use periodic boundary condition in our calculations, and the system size is fixed as $4096\times 4096$ for
single-layer BP, and $2048\times 2048$ for bilayer. 

\section{Model of Defects}
In the employed TB model, the point defects are modeled by elimination of
atoms randomly over the whole sample, which can be viewed as 
phosphorus single vacancies, chemical adsorbates such as covalently bonded
adatoms or admolecules, or  substitution of other types of atoms which
prevent the electronic hopping to their neighbors\cite%
{Neto2009,Sarma2011,Katsnelsonbook,WK10,YRK10,YRK10b,QiuH2013,Yuan2014}. The
amount of point defects is described by $n_{x}$, which is the probability
for a single defect to appear at one lattice site. The electron-hole
puddles, in which the spatial charge inhomogeneity leads to a local change
of on-site potentials, can be represented as a correlated Gaussian potential
in the TB model\cite{LMC08,YRK10b,YRRK11}. The value of the potential at
site $i $ follows 
% $v_{i}=\sum_{k=1}^{N_{imp}^{v}}U_{k}\exp \left(
% -\left\vert \mathbf{r}_{i}-\mathbf{r}_{k}\right\vert ^{2}/2d^{2}\right) $, 
\begin{equation}
v_{i}=\sum_{k=1}^{N_{imp}^{v}}U_{k}\exp \left( -\frac{\left\vert \mathbf{r}%
_{i}-\mathbf{r}_{k}\right\vert ^{2}}{2d^{2}}\right) ,  \label{vgaussian}
\end{equation}%
where $N_{imp}^{v}$ is the number of the Gaussian centers, $\mathbf{r}_{k}$
is the position of the $k-th$ Gaussian center, which are chosen to be
randomly distributed over the centers of the projected lattice on the
surface, $U_{k}$ is the amplitude of the potential at the Gaussian center,
which is uniformly random in the range $[-\Delta ,\Delta ]$, and $d$ is
interpreted as the effective potential radius. The typical values of $d$
used in our model are $\Delta =5eV$ and $d=5a$ for electron-hole puddles\cite%
{YRK10b,YRRK11}. 
% , where $a\approx 2.216$\r{A}\ is the atomic distance of two
% nearest neighbors within the same plane. 
Similarly, the amount of electron-hole puddles is measured by $n_{c}$, which
is the probability for a Gaussian potential to appear.

\begin{figure}[t]
\begin{center}
\mbox{
\includegraphics[width=4.25cm]{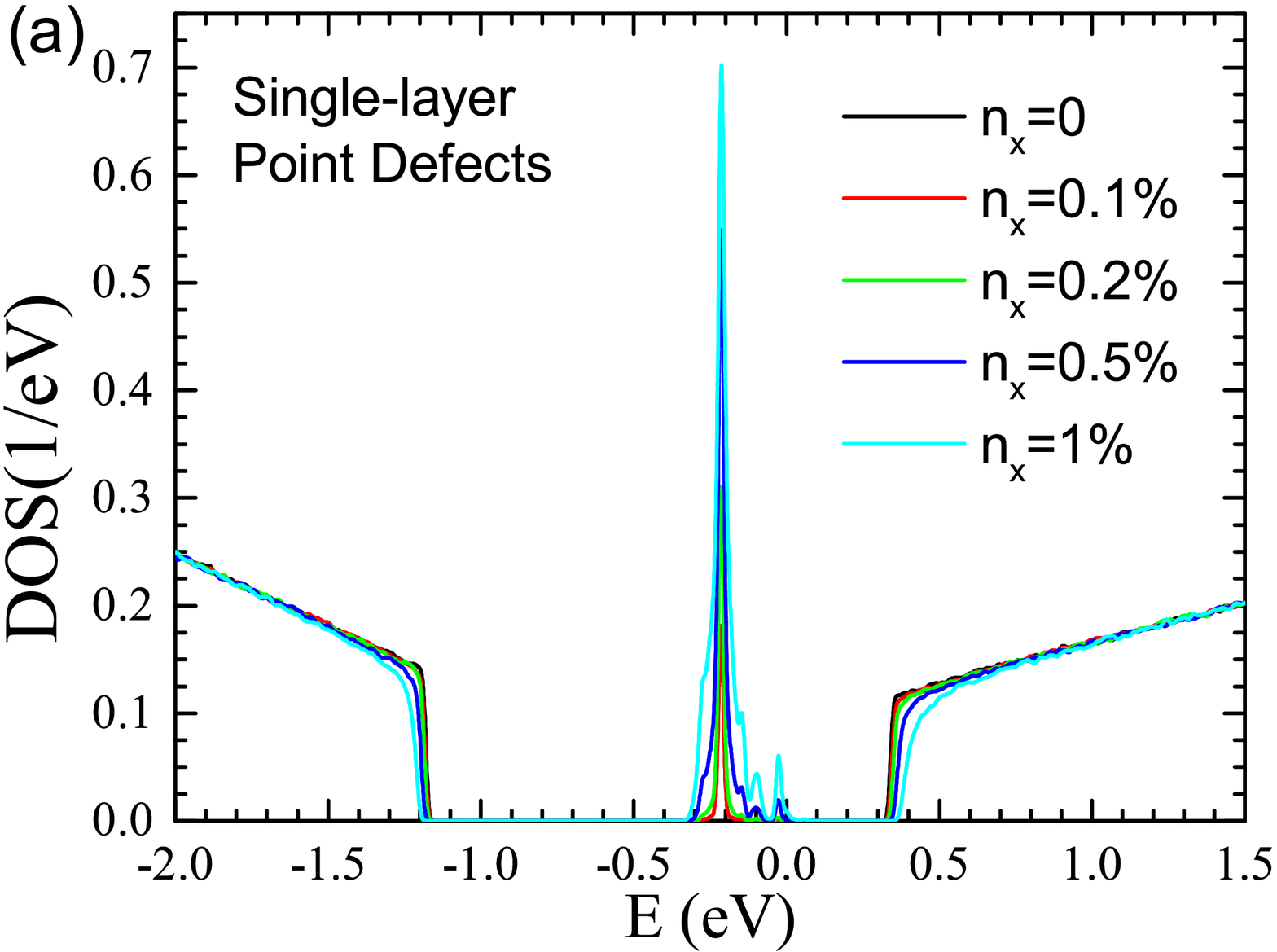}
\includegraphics[width=4.25cm]{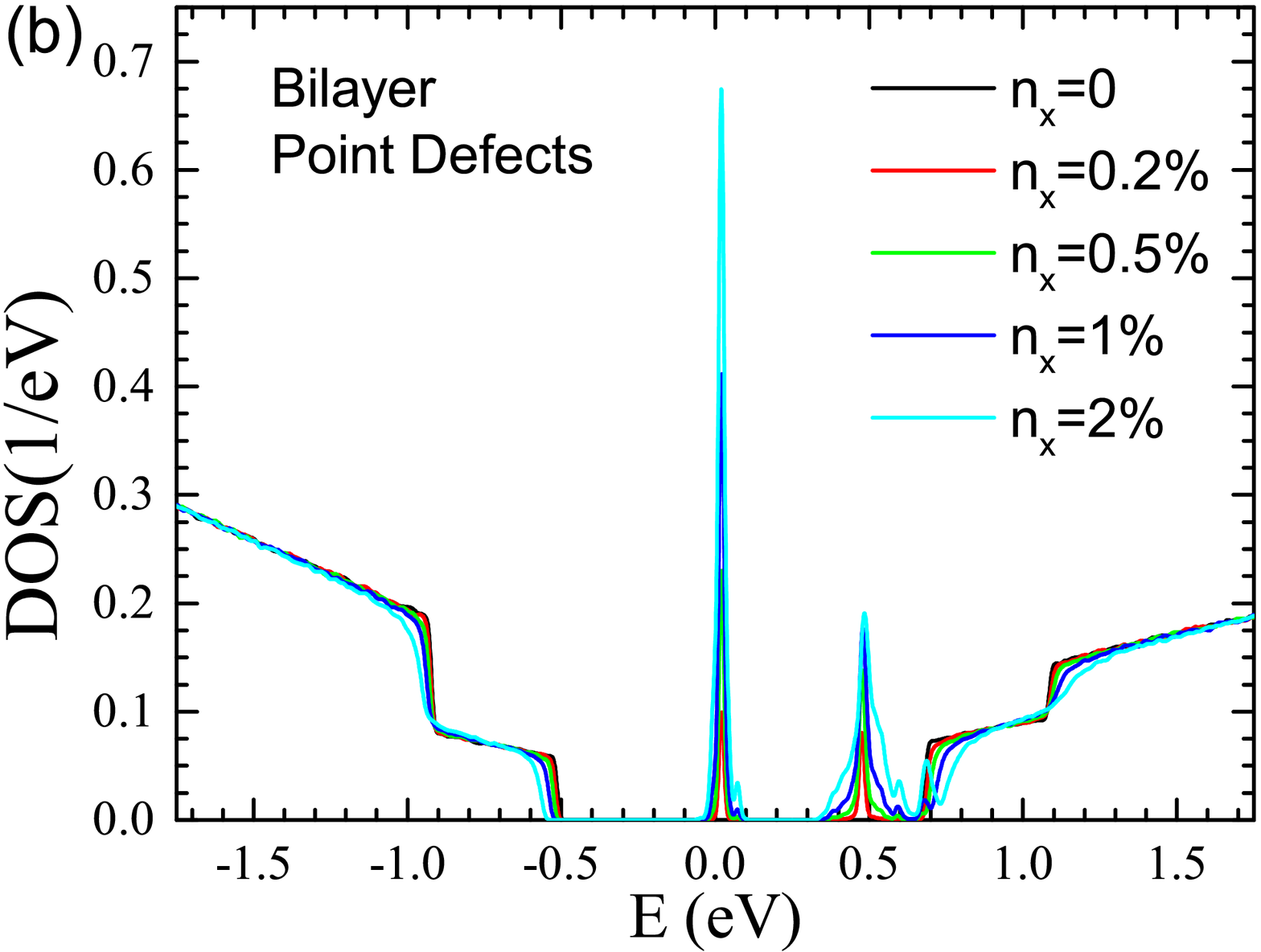}
} 
\mbox{
\includegraphics[width=4.25cm]{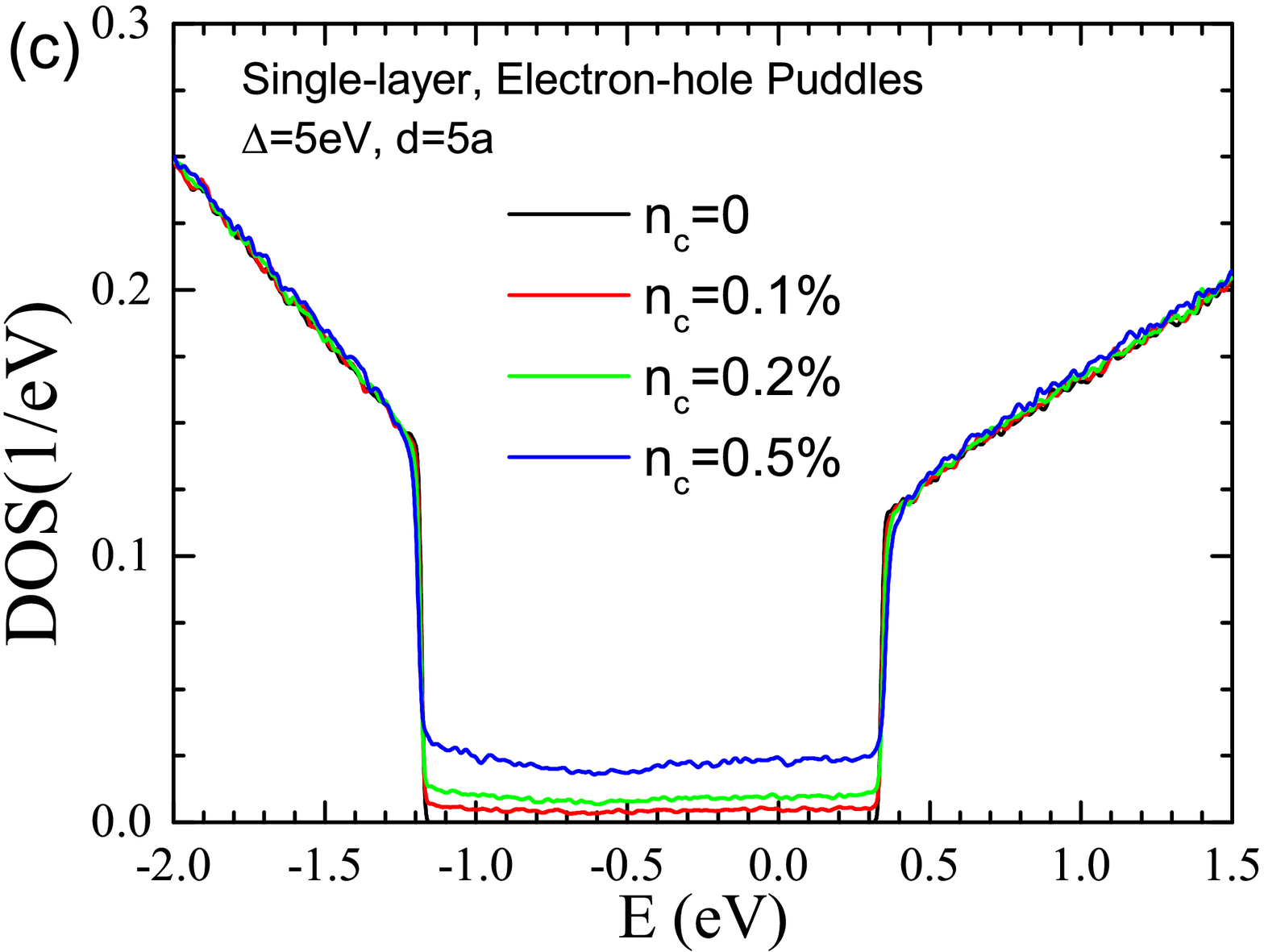}
\includegraphics[width=4.25cm]{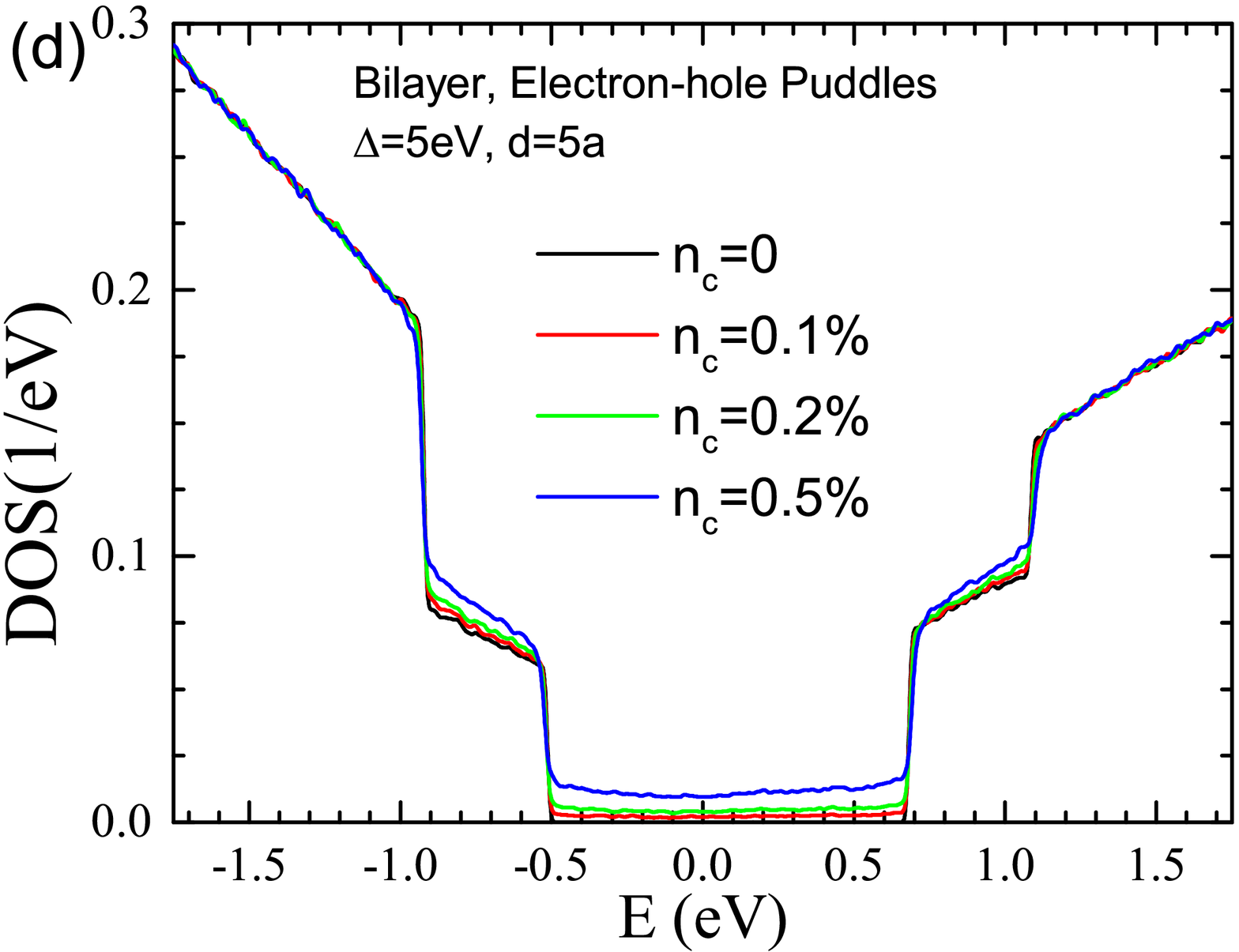}
}
\end{center}
\caption{Density of states of single-layer (left columns) and bilayer (right
columns) with point defects (upper rows) or electron-hole puddles (bottom
rows). The value $n_{x}(n_{c})=0.1\%$ corresponds to defect concentration as 
$2.98(1.49)\times 10^{12}$ per cm$^{2}$.}
\label{Fig:DOS}
\end{figure}

\begin{figure*}[t]
\begin{center}
\mbox{
\includegraphics[width=4.25cm]{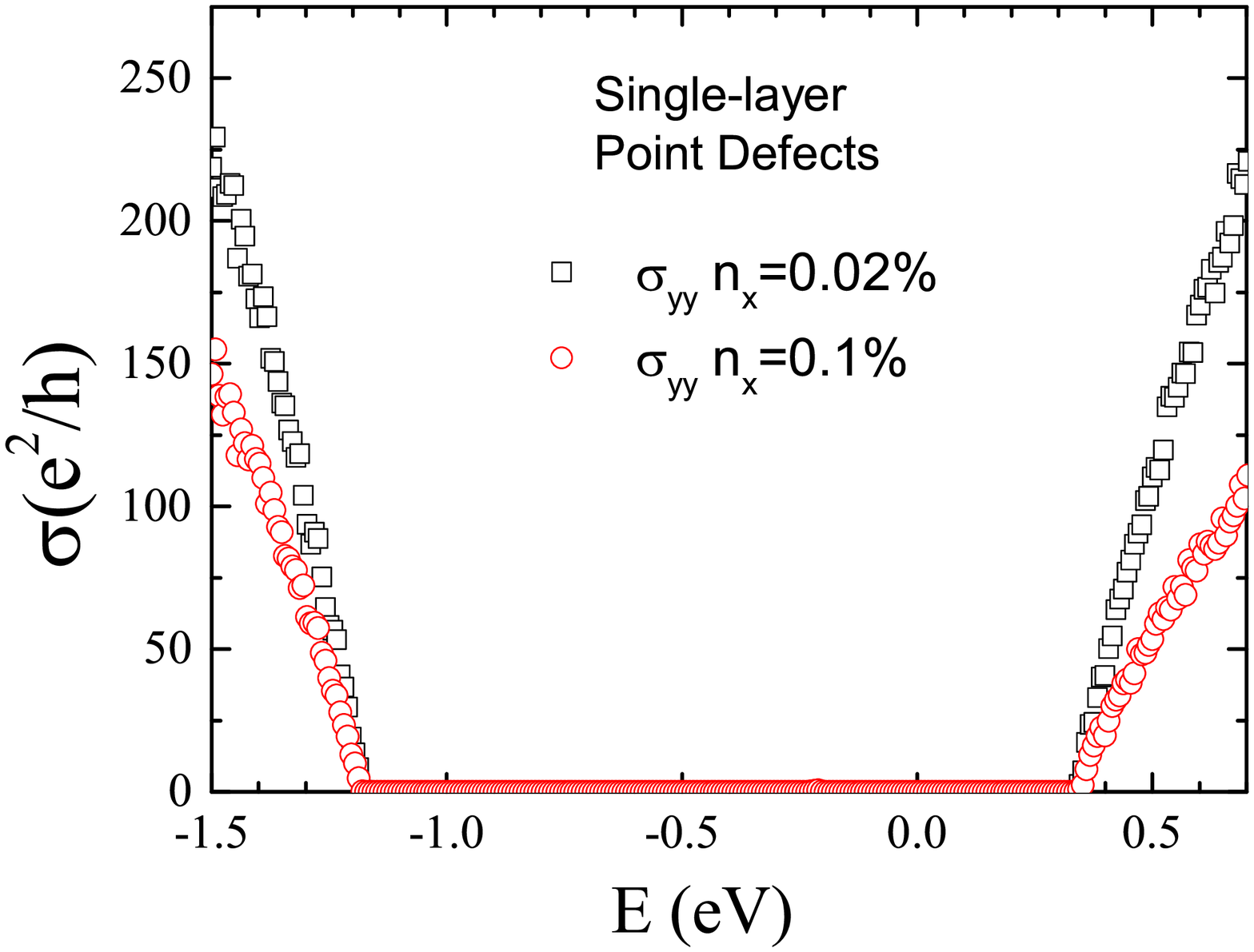}
\includegraphics[width=4.25cm]{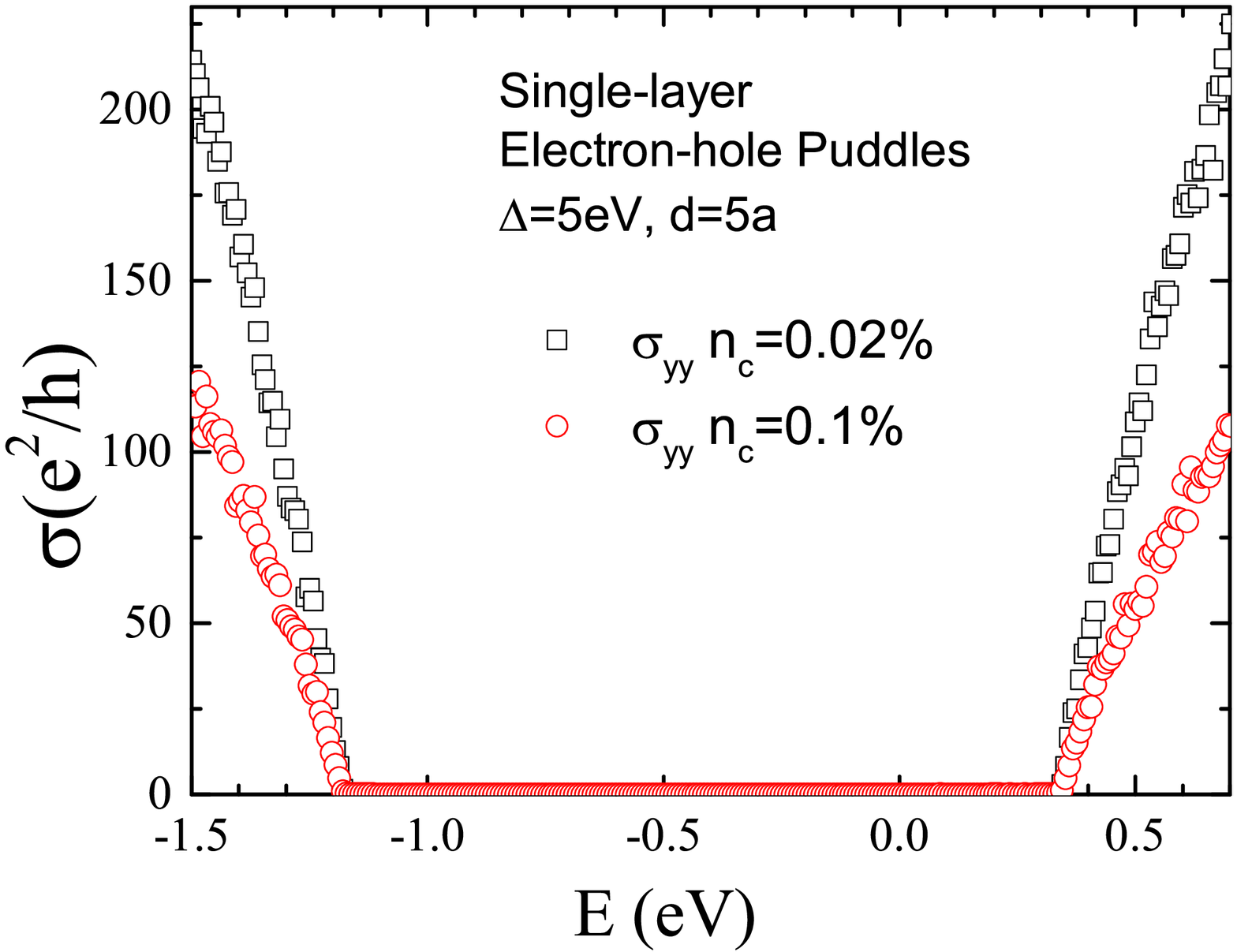}
\includegraphics[width=4.25cm]{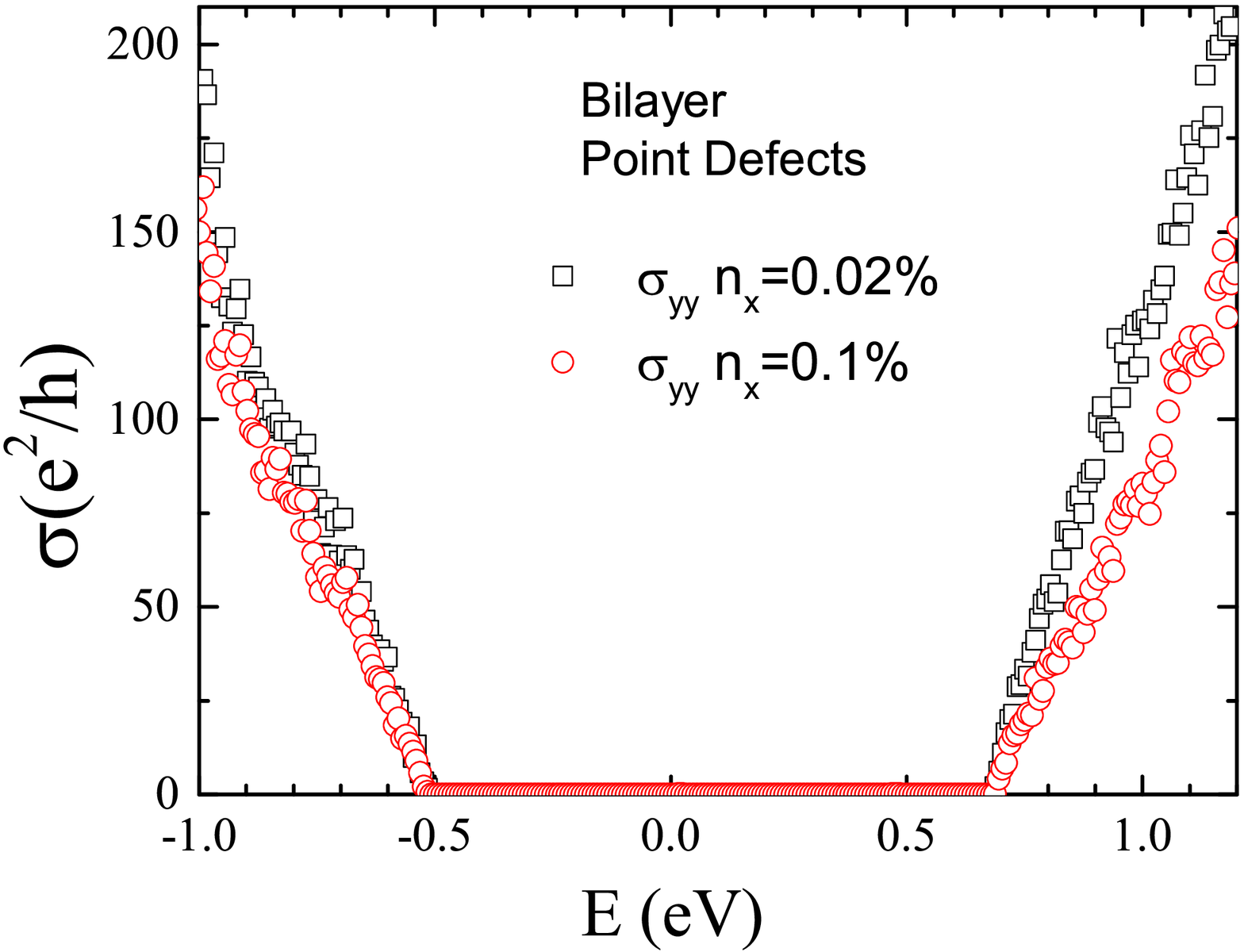}
\includegraphics[width=4.25cm]{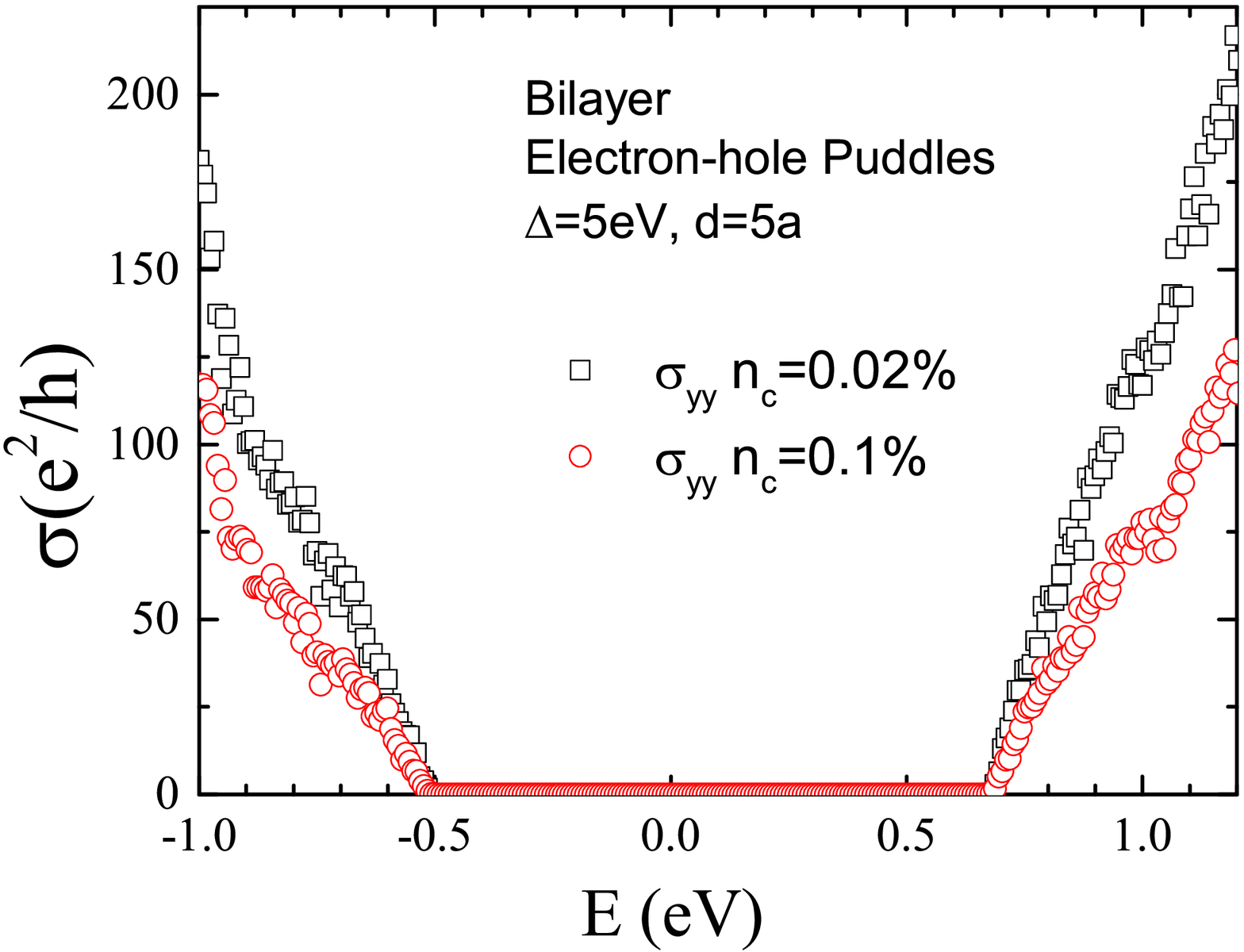}
} 
\mbox{
\includegraphics[width=4.25cm]{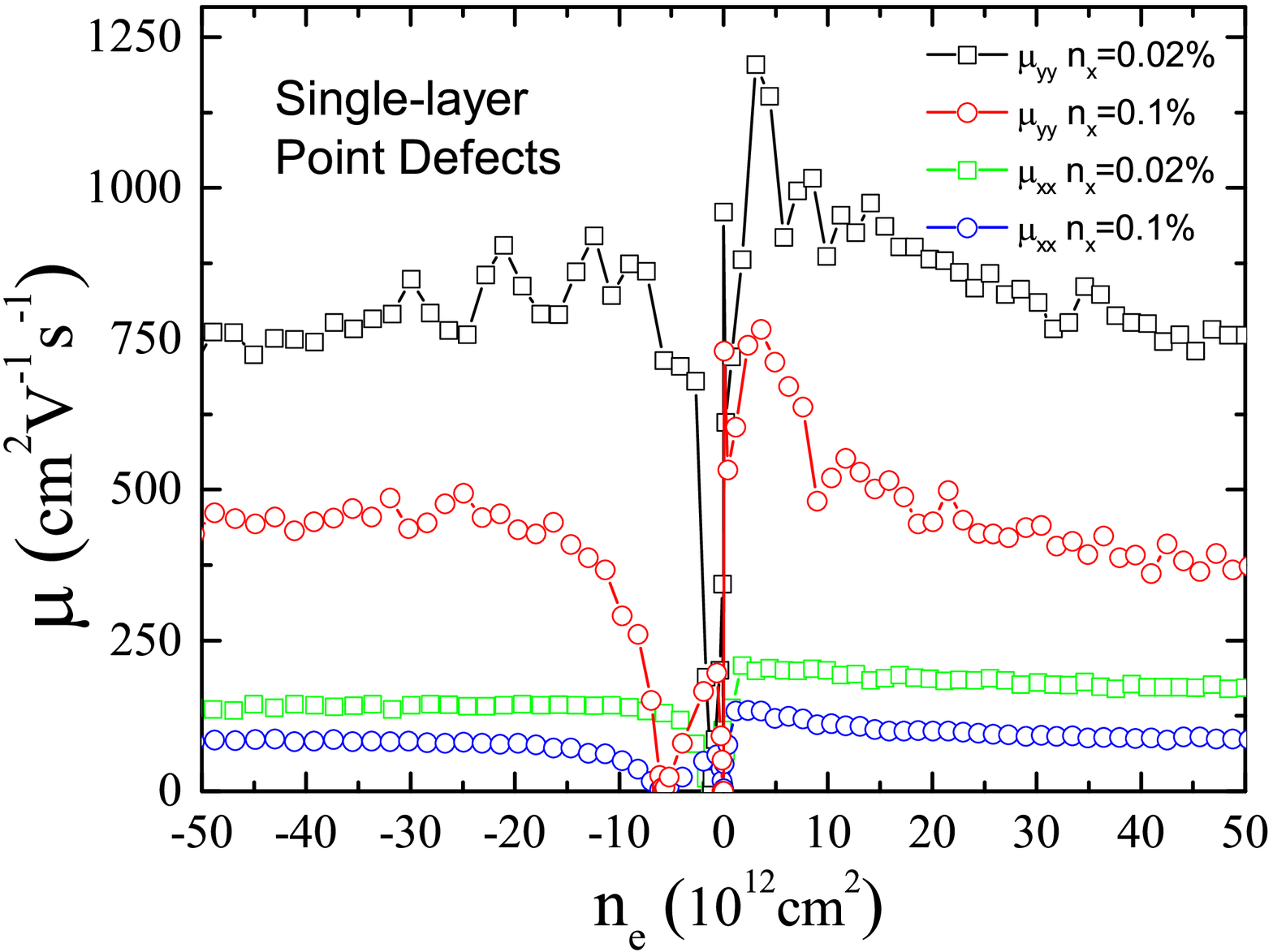}
\includegraphics[width=4.25cm]{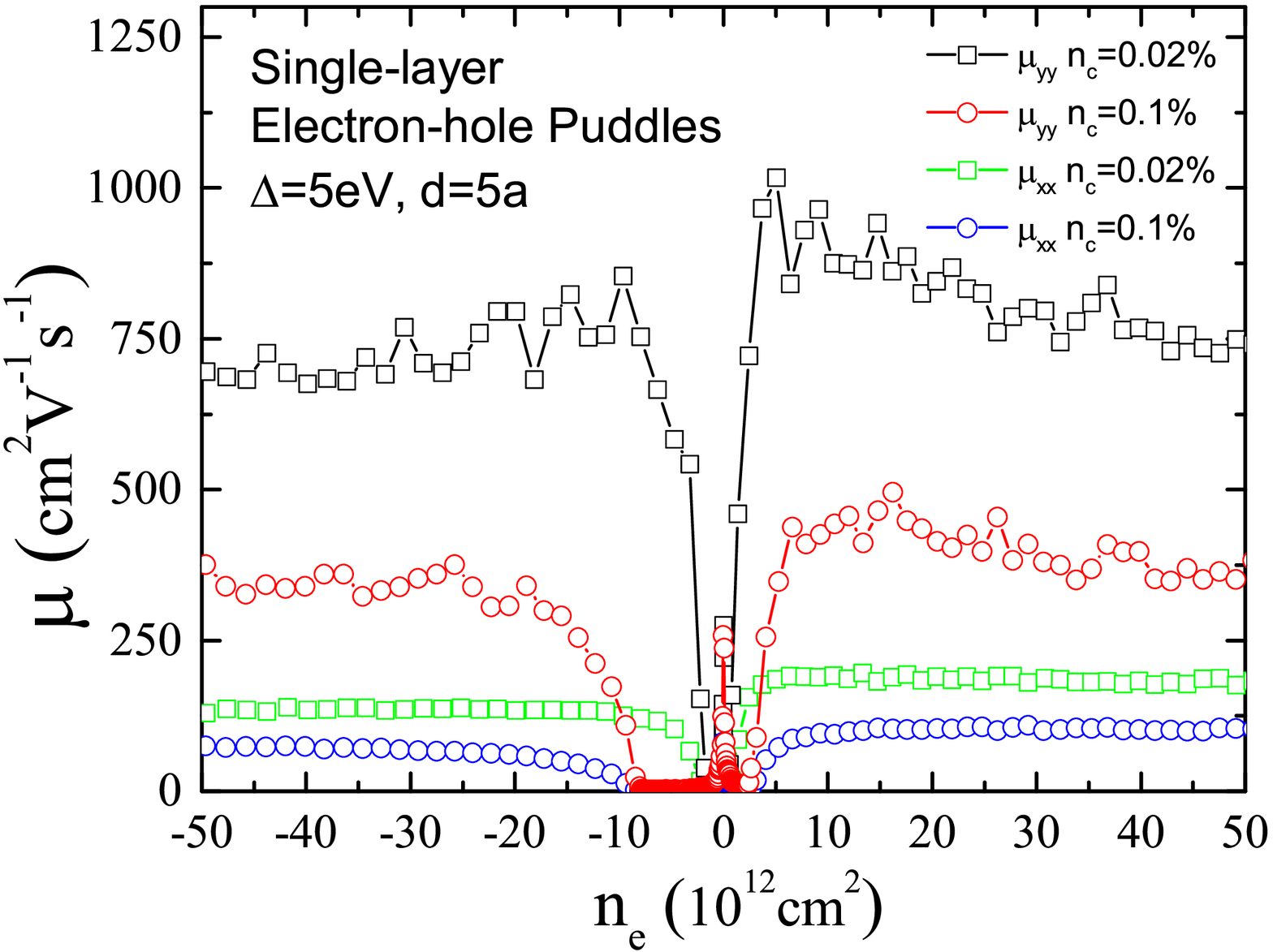}
\includegraphics[width=4.25cm]{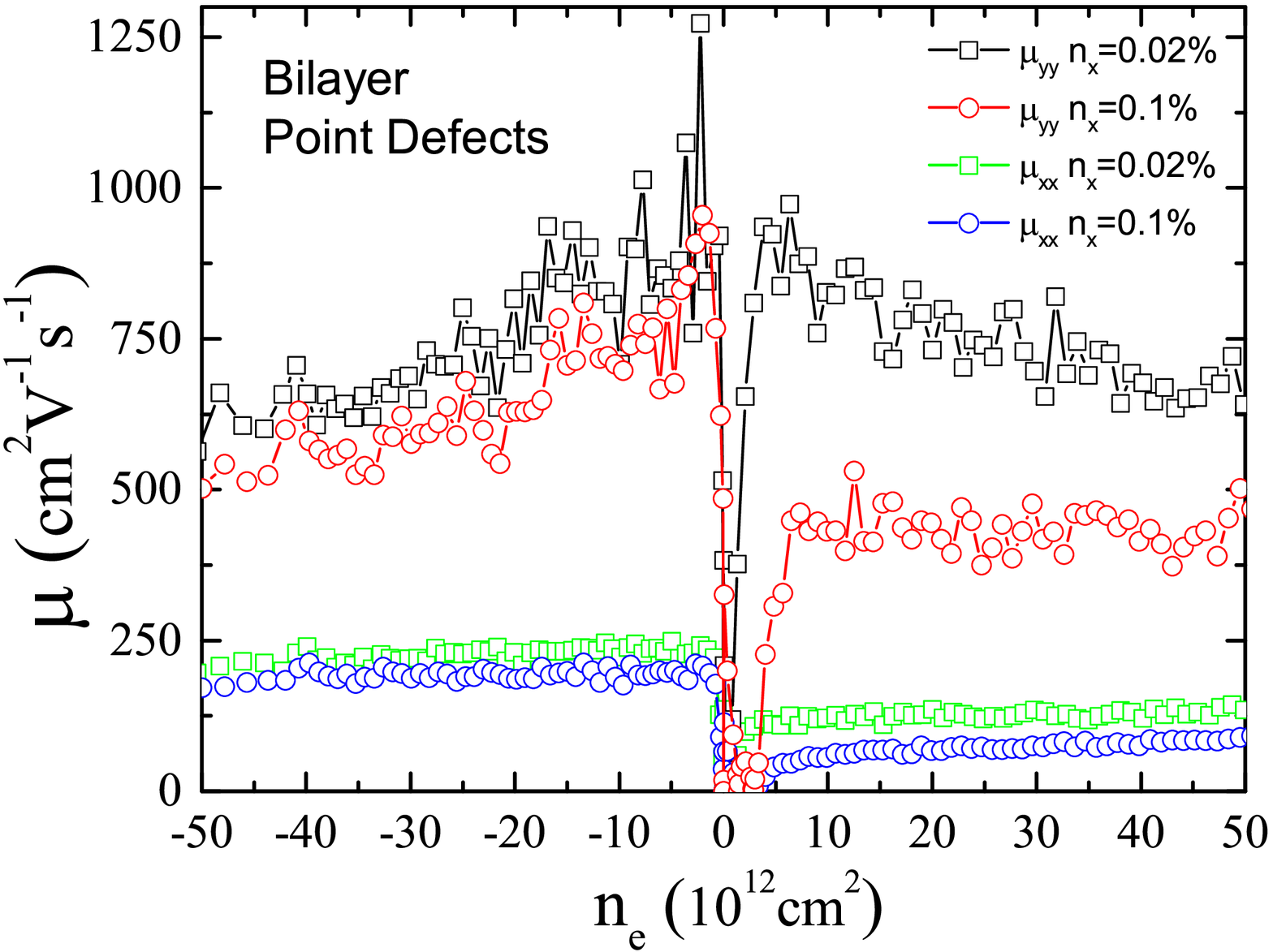}
\includegraphics[width=4.25cm]{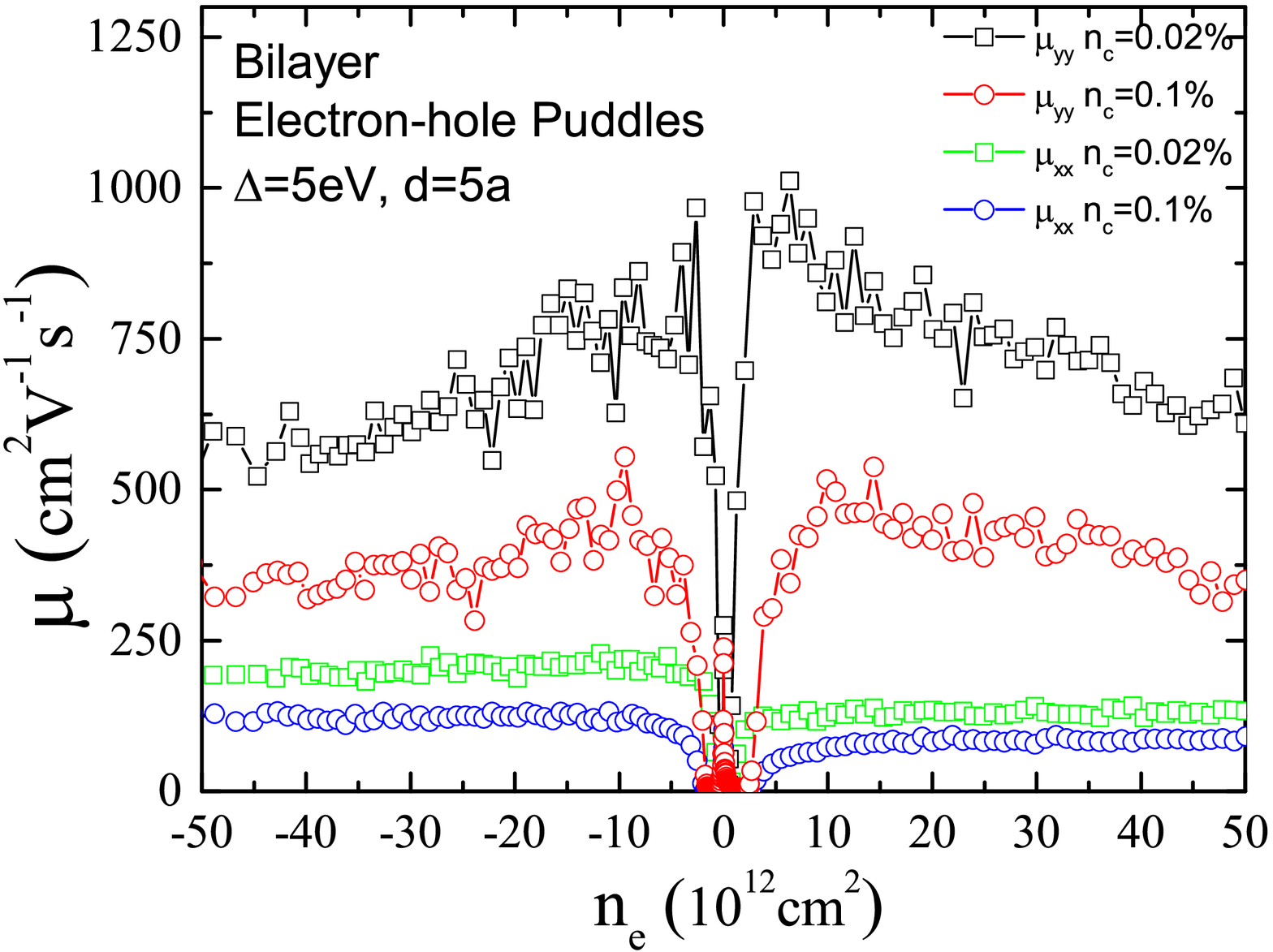}
}
\end{center}
\caption{Transport properties of single- and bilayer with defects. Top: dc conductivity as a function of doping energy;
Bottom: carrier mobility as a function of carrier density.}
\label{Fig:DC}
\end{figure*}

In order to investigate the influence of the defects on the electronic
structure of BP, we calculate the density of states (DOS) of samples with
randomly distributed point defects or Gaussian potentials. As we expect, the
presence of point defects leads to midgap states within the energy band gap,
which can be identified by the sharp peaks in the DOS shown in Fig.~\ref{Fig:DOS}(a)
and (b), where the number of the defect states is proportional to the
concentration of point defects\cite{YRK10,Yuan2014}. Similar midgap states appear
in the first-principles calculations of BP with different types of point defects\cite{Liu2014,HuW2014,Ziletti2015,Zhang2014,Kulish2015}.
On the other hand, the presence of electron-hole puddles does not introduce any resonant states
[see Fig.~\ref{Fig:DOS}(c) and (d)]. Instead, there is a uniform enhancement
of the DOS within the gap due to the random distribution of positive and
negative potentials, whereas the increased amplitude is proportional to the
number (concentration) of potential puddles.

\section{Transport Properties}

The calculations of transport properties by using the Kubo formula
show that for both point defects and electron-hole puddles, the impurity states
within the band gap are insulating states (see
Fig. \ref{Fig:DC}). This is due to the Anderson localization in disordered 2D
system\cite{Ziman1979,Mott2012}, and the result is not sensitive to the defect
concentration. On the other hand, the dc conductivities in the valence and
conduction bands decrease monotonically as the defect concentration
increases. This is consistent with the qualitative analysis by using the $T$ matrix.\cite%
{WKL09,Yuan2014}. For example, point defects in our model is represented
as impurities with infinite on-site potential, and the scattering by these impurities 
leads to $T\left( E\right) \rightarrow -1/g_{0}(E)$, where $%
g_{0}(E)$ is the local unperturbed Green's function. For a semiconductor
with electron-hole asymmetry like BP, the density of states follows
approximately $N_{0}(E)=D_{c}\Theta (E-E_{c})+D_{v}\Theta (E_{v}-E)$, where $%
E_{c(v)}$ is the energy at the edge of conduction (valence) band. The local
unperturbed Green's function follows\cite{WKL09,Yuan2014}%
\begin{equation}
g_{0}(E)=D_{c}\log \left\vert \frac{E-E_{c}}{E-W_{c}}\right\vert +D_{v}\log
\left\vert \frac{E+W_{v}}{E-E_{v}}\right\vert
\end{equation}%
where $W_{c(v)}$ is the width of the conduction (valence) bands. The dc
conductivity is proportional to the inverse of the defect concentration $n_{x}$,
as it can be expressed as $\sigma =(2e^{2}/h)E\tau $ where $\tau
^{-1}=(2\pi /\hbar )n_{x}|T(E)|^{2}N_{0}(E)$ is the scattering rate in terms of $n_{x}$.

% The calculations of transport properties by using the Kubo formula show that
% for both point defects and electron-hole puddles, the defect states within
% the band gap are insulating states (see Supplementary Materials for more
% details). This is due to the Anderson localization in disordered 2D system%
% \cite{Ziman1979,Mott2012}, and the result is not sensitive to the defect
% concentration. The anisotropic carrier mobility plotted in Fig.~\ref{Fig:DC} shows clearly
% that the mobility along the armchair direction is much higher than the zigzag direction. 
Furthermore, the carrier-density dependence of the mobility
shows different electron-hole asymmetry in single- and bilayer.
The electron mobility is higher than the hole one in
single-layer, but it is opposite in bilayer. For example, for single-layer with
defect concentration $n_{x}=0.02\%$, the electron mobility along the armchair
direction is about $1200$ $cm^{2}V^{-1}s^{-1}$ at carrier density $%
n_{e}=5\times 10^{12}cm^{2}$, which is larger than the hole mobility ($700$ $%
cm^{2}V^{-1}s^{-1}$) at the same order of carrier density. 
The asymmetry of the mobility becomes more obvious with the
increased number of defects. For the same carrier density considered above,
when the defect concentration increases to $n_{x}=0.1\%$, the mobility drops
to $700$ $cm^{2}V^{-1}s^{-1}$ for electrons and less than $50$ $%
cm^{2}V^{-1}s^{-1}$ for holes. On the contrary, for bilayer with $%
n_{x}=0.1\%$, the electron mobility is about $300$ $cm^{2}V^{-1}s^{-1}$ at
carrier density $n_{e}=5\times 10^{12}cm^{2}$, much smaller than the
corresponding hole mobility ($800$ $cm^{2}V^{-1}s^{-1}$). The different
electron and hole mobility suggests that single-layer is more suitable
for the application as an $n$-doped field-effect transistor, while bilayer
is better as a $p$-doped field-effect transistor. The drain current
modulation of $n$-doped single-layer and $p$-doped bilayer can reach
the experimental observed value ($\sim 10^{5}$ in {Ref.~\onlinecite{LLi2014})%
} even with defect concentration $n_{x}(n_{c})=0.1\%$.

\section{Optical Properties}

\begin{figure*}[t]
\begin{center}
\mbox{
\includegraphics[width=4.25cm]{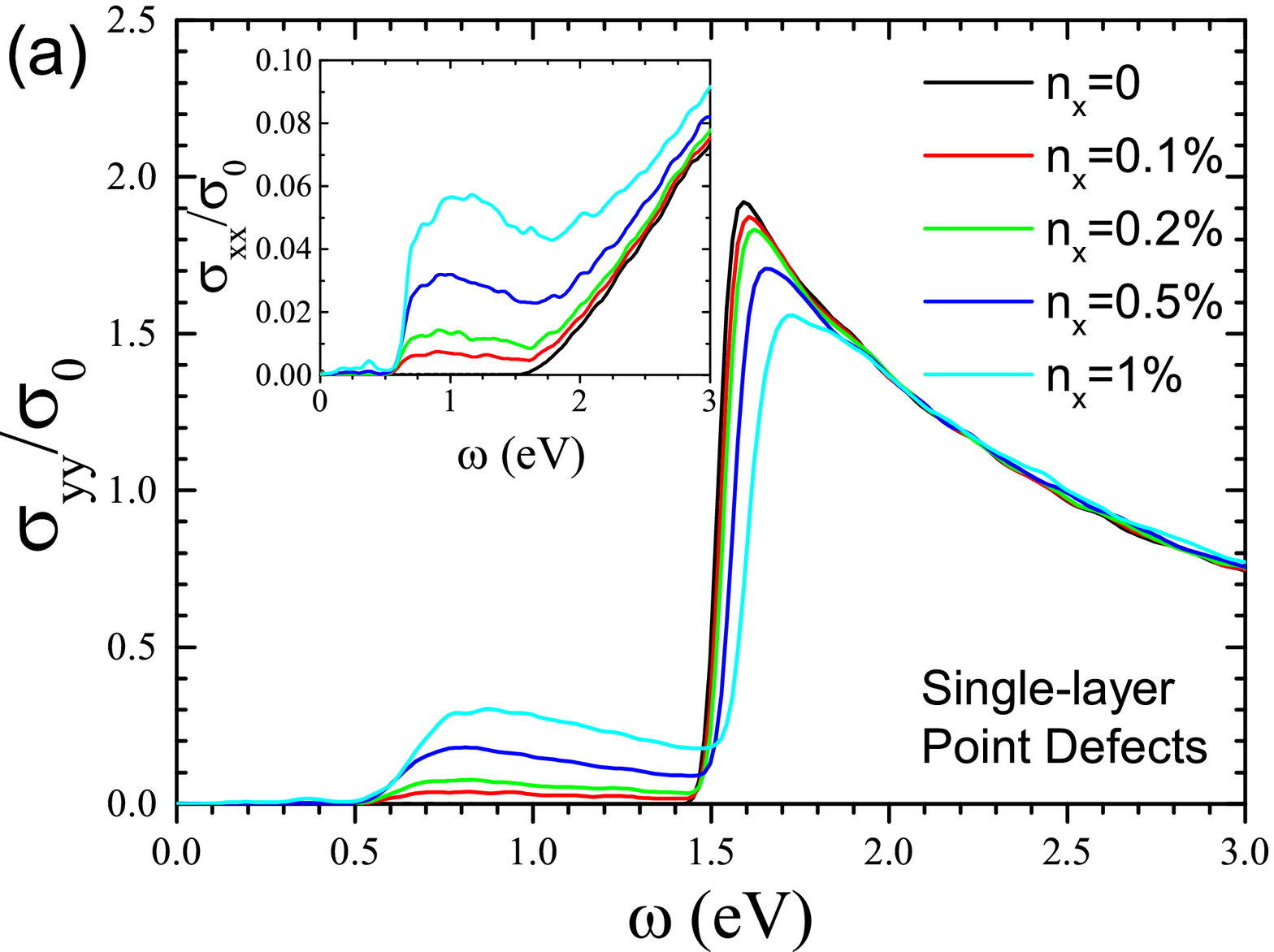}
\includegraphics[width=4.25cm]{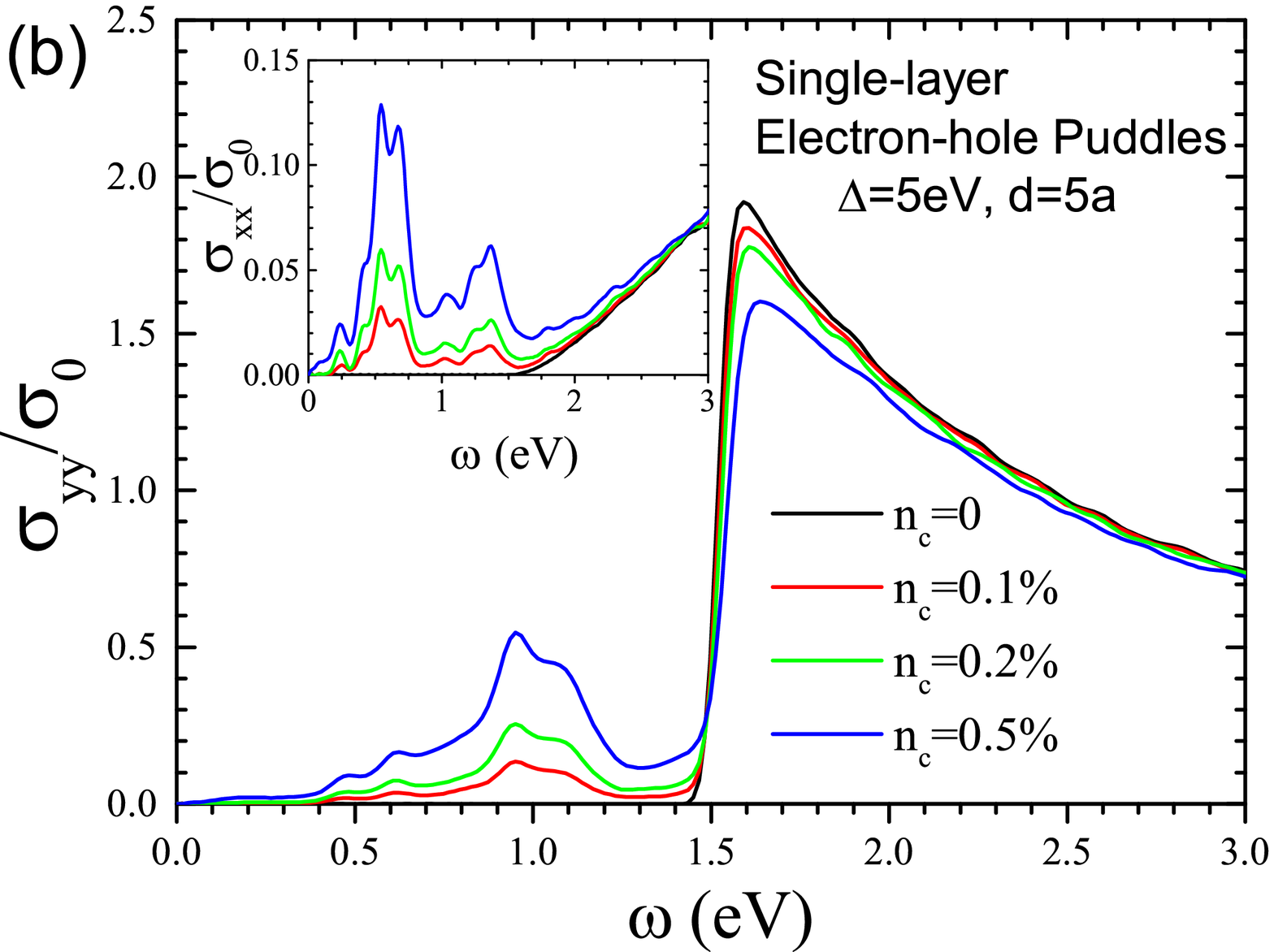}
\includegraphics[width=4.25cm]{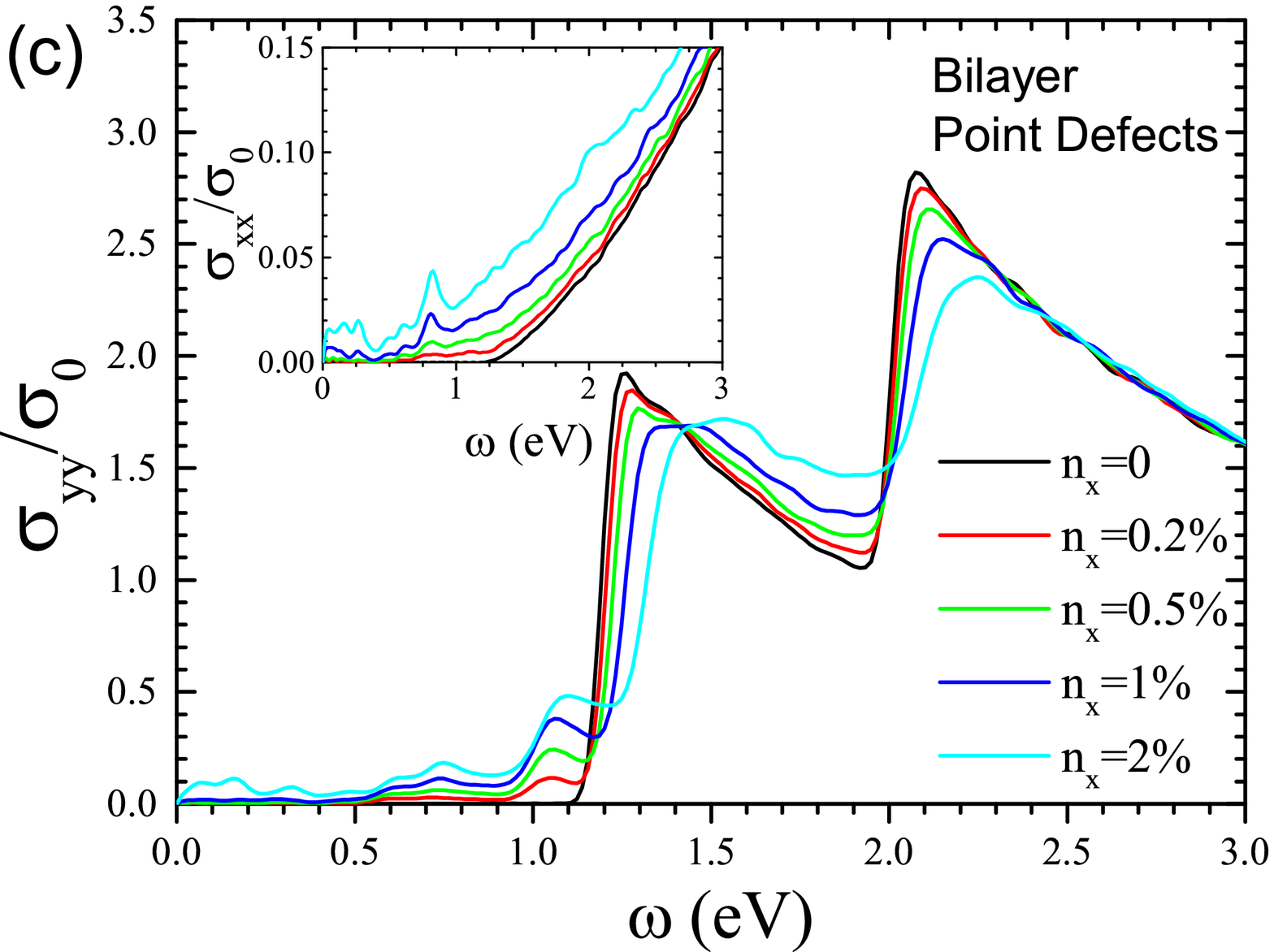}
\includegraphics[width=4.25cm]{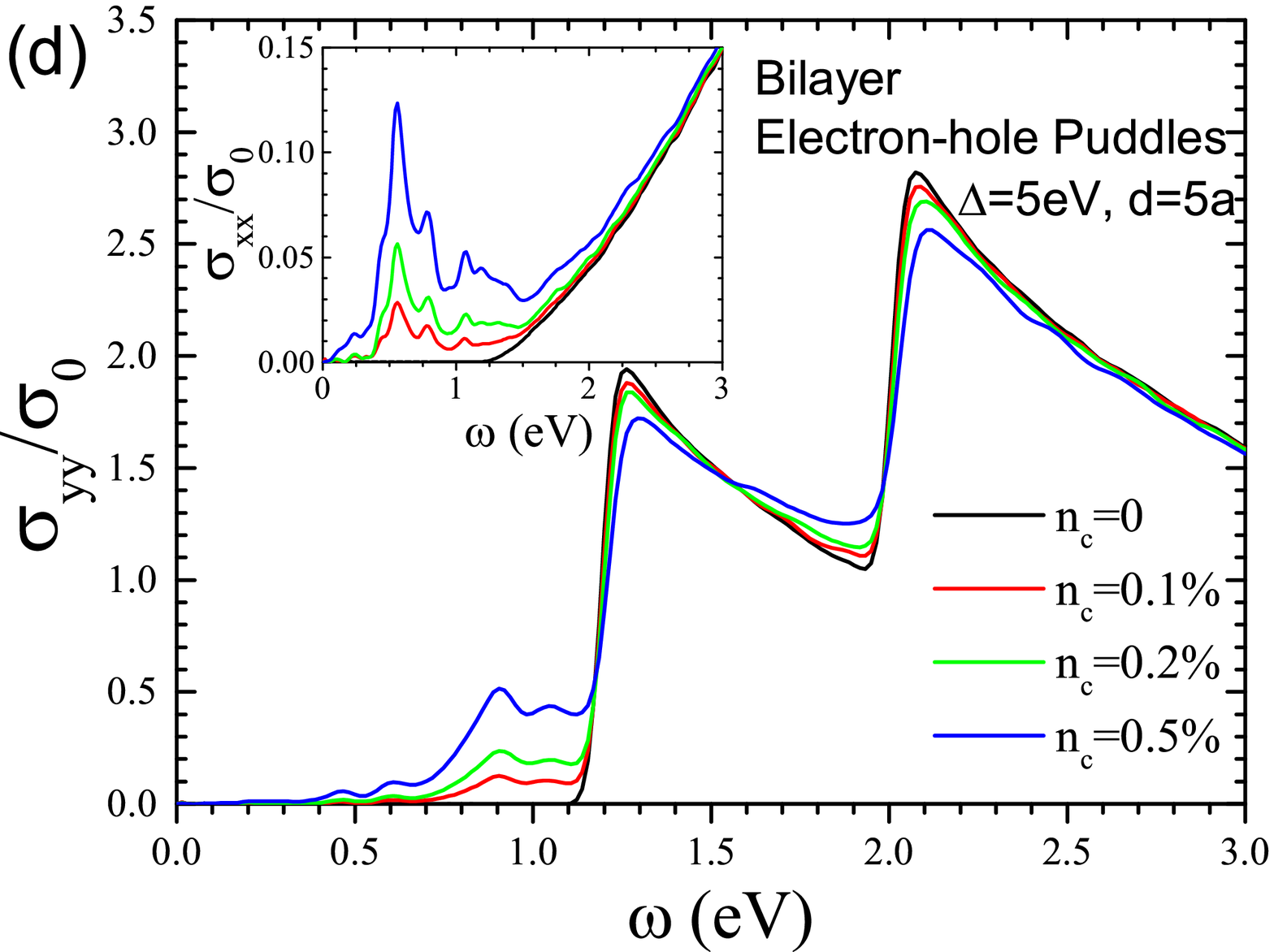}
} 
\mbox{
\includegraphics[width=4.25cm]{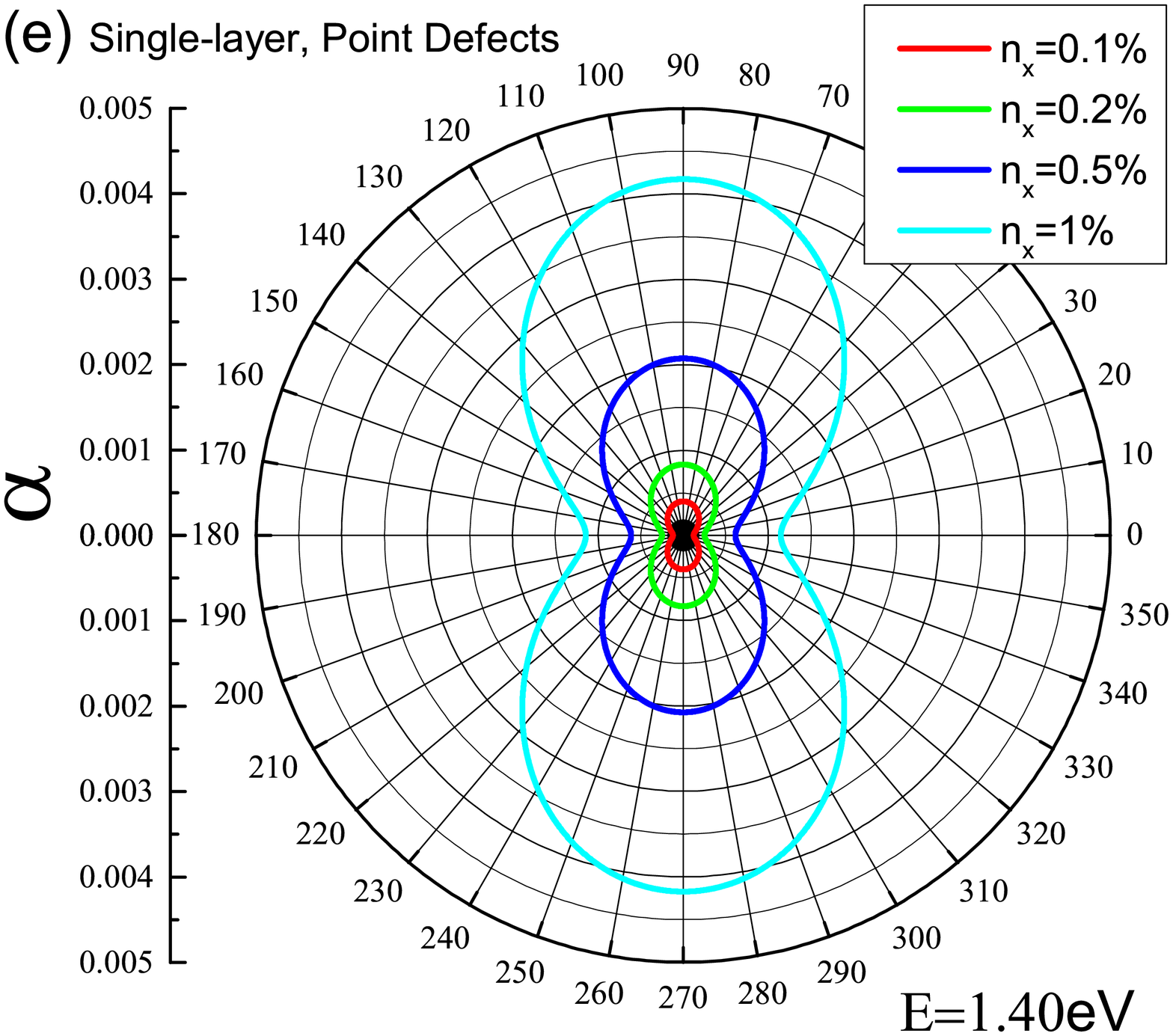}
\includegraphics[width=4.25cm]{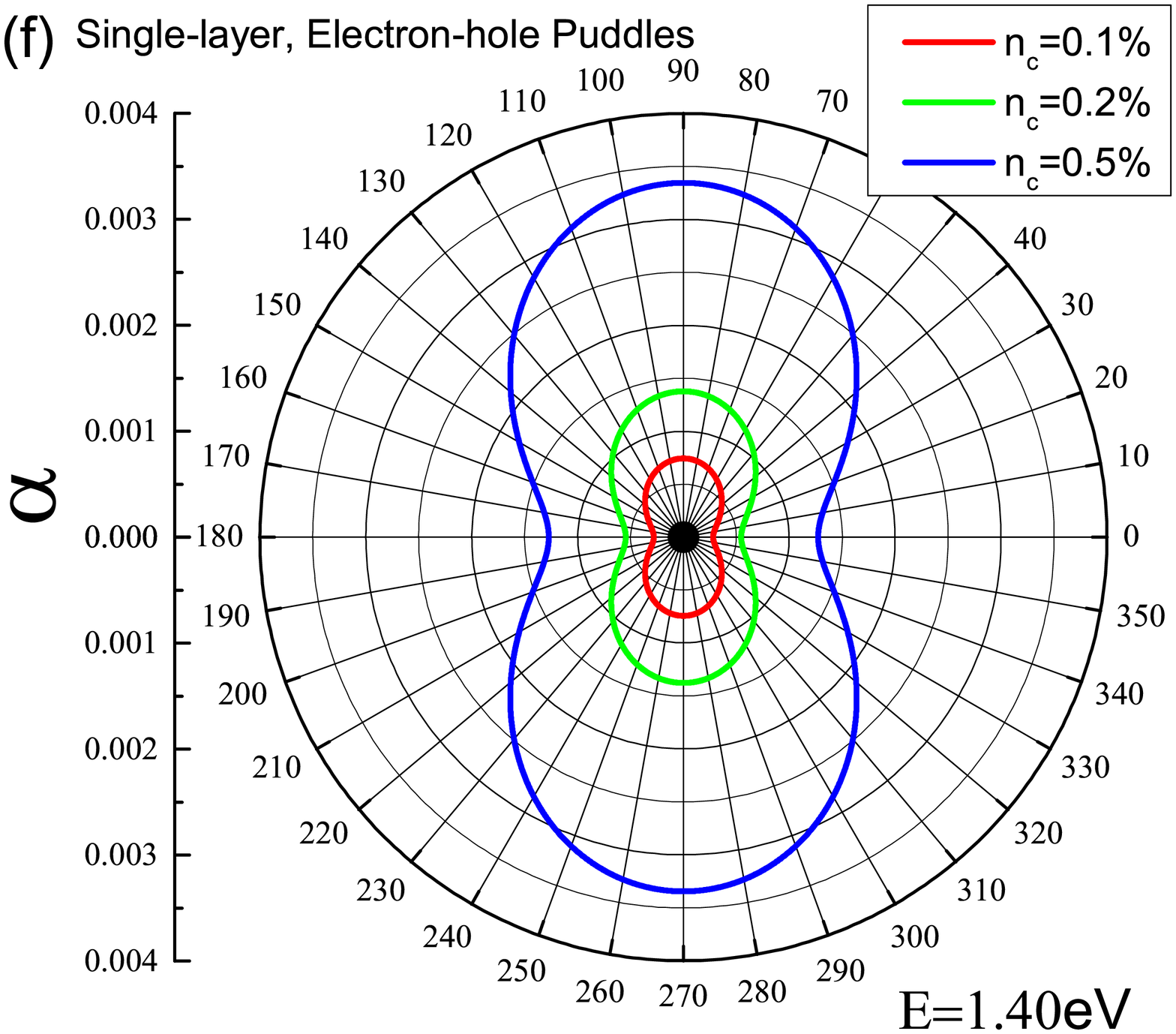}
\includegraphics[width=4.25cm]{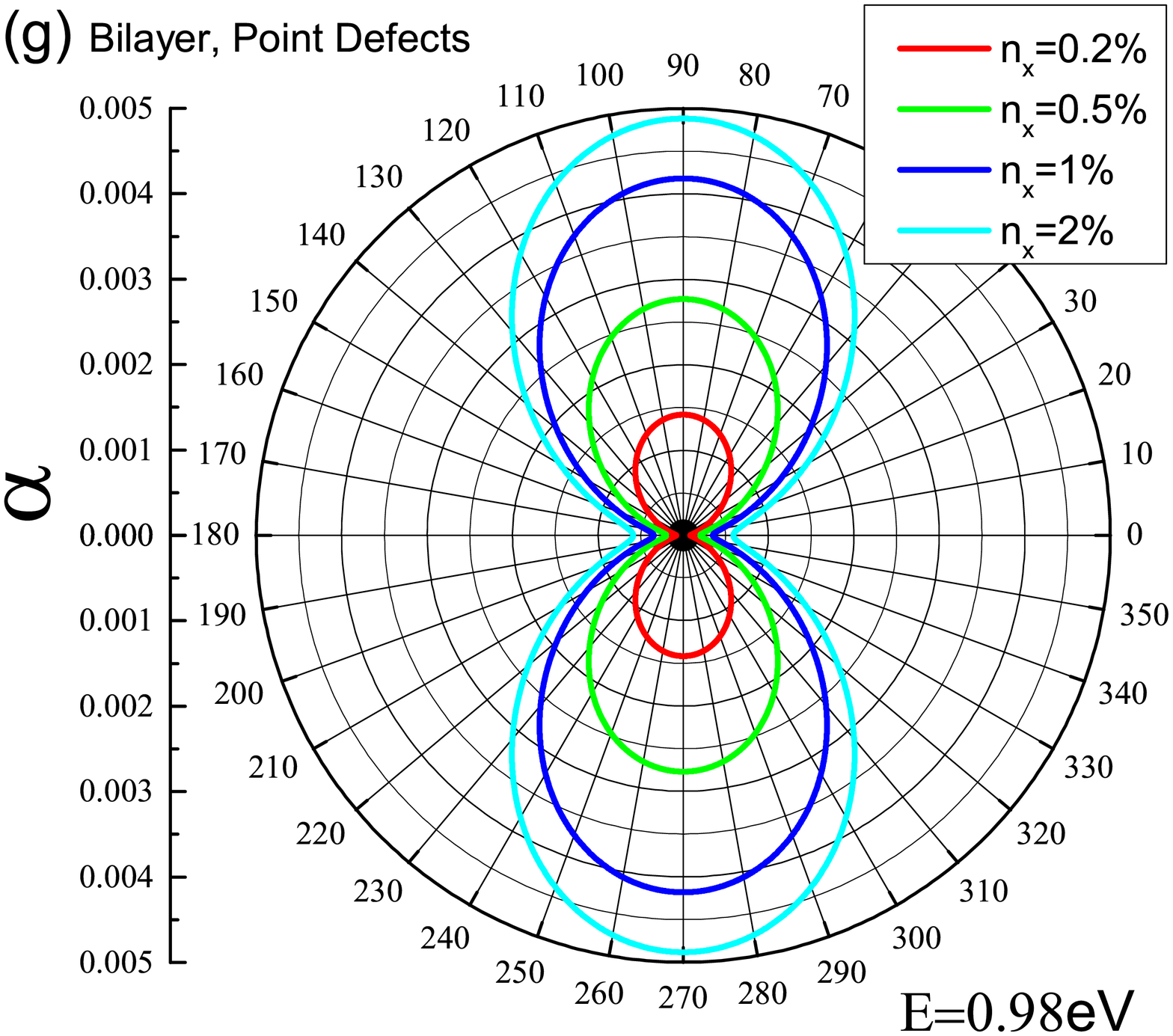}
\includegraphics[width=4.25cm]{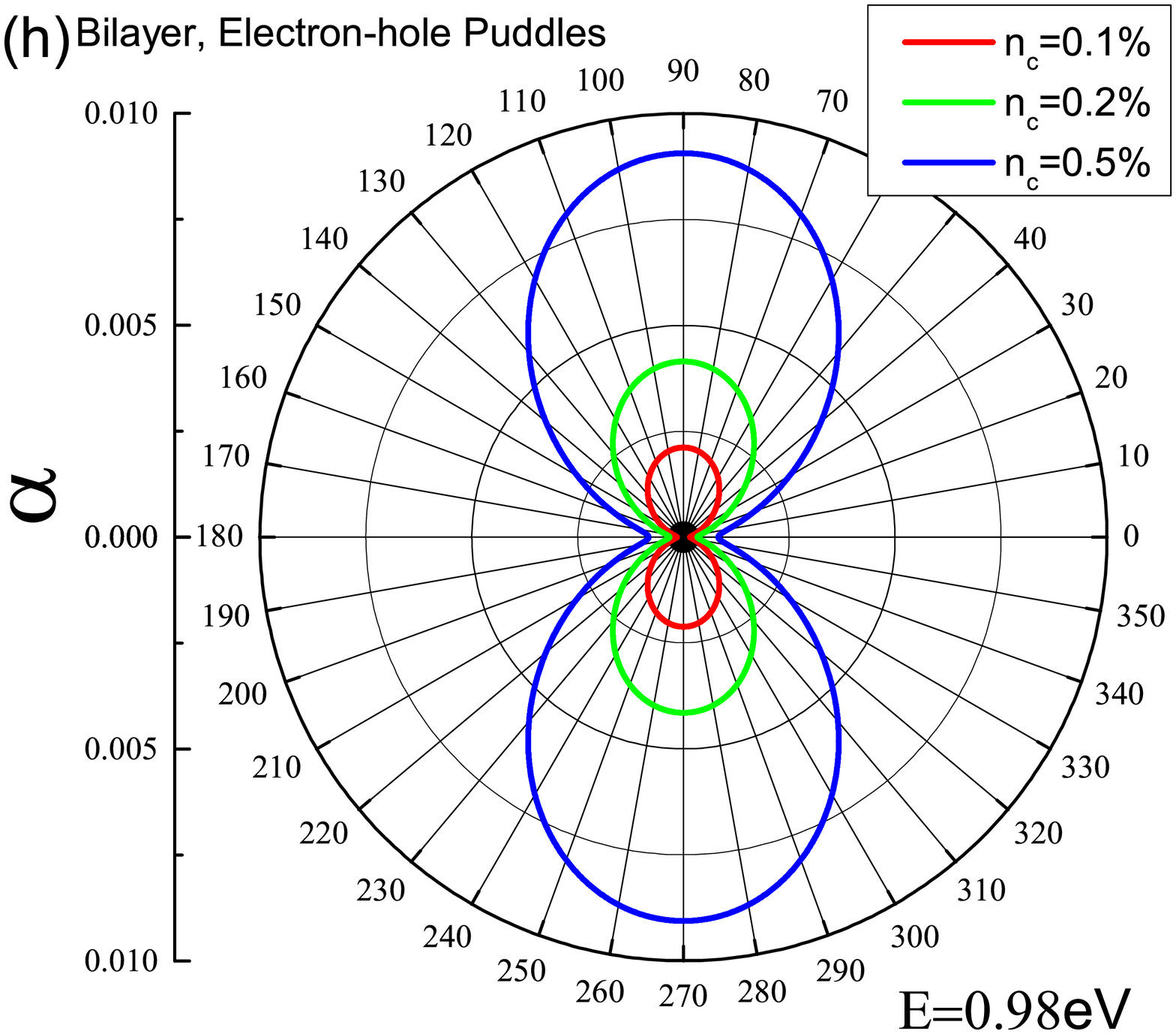}
} 
\mbox{
\includegraphics[width=4.25cm]{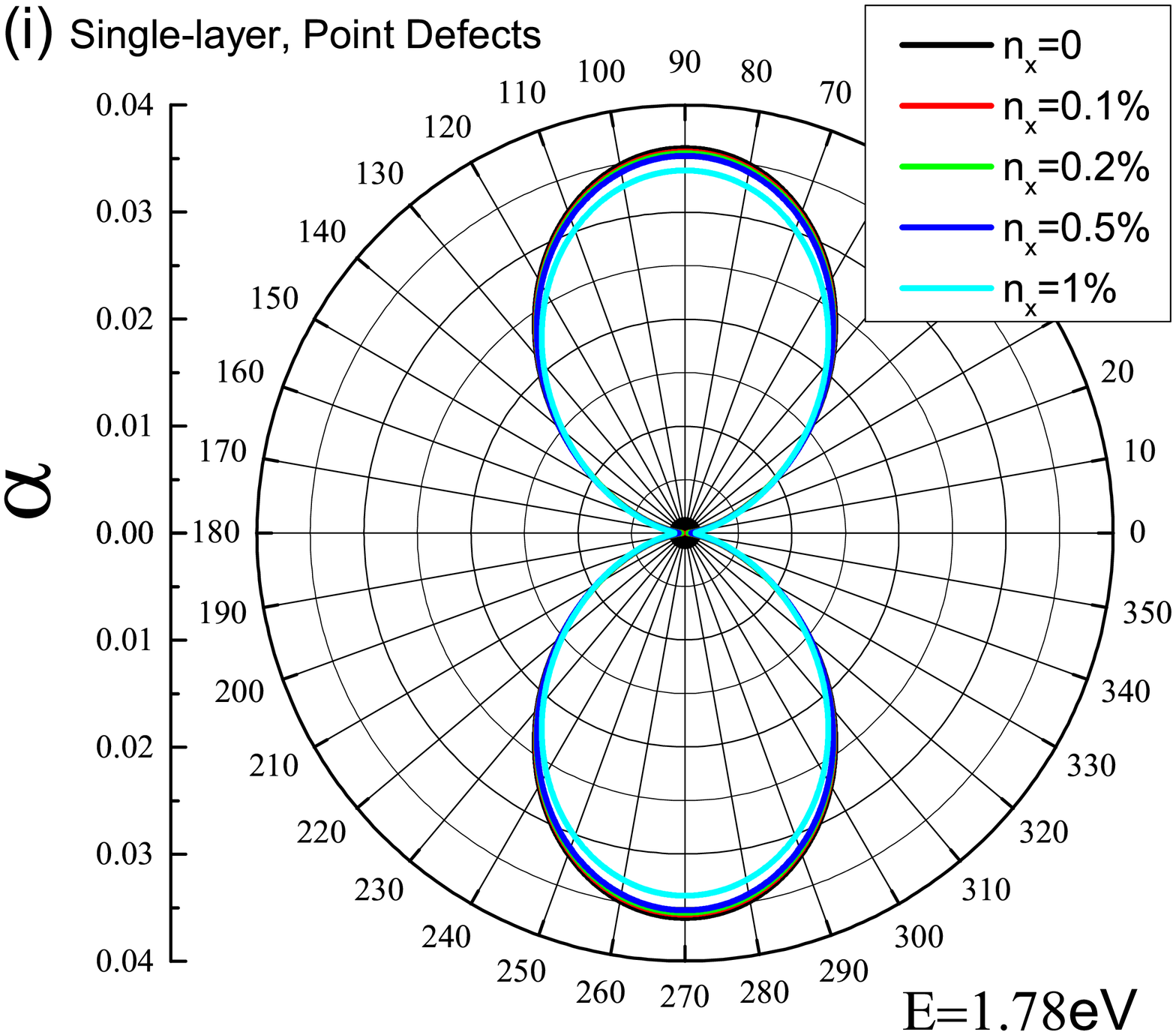}
\includegraphics[width=4.25cm]{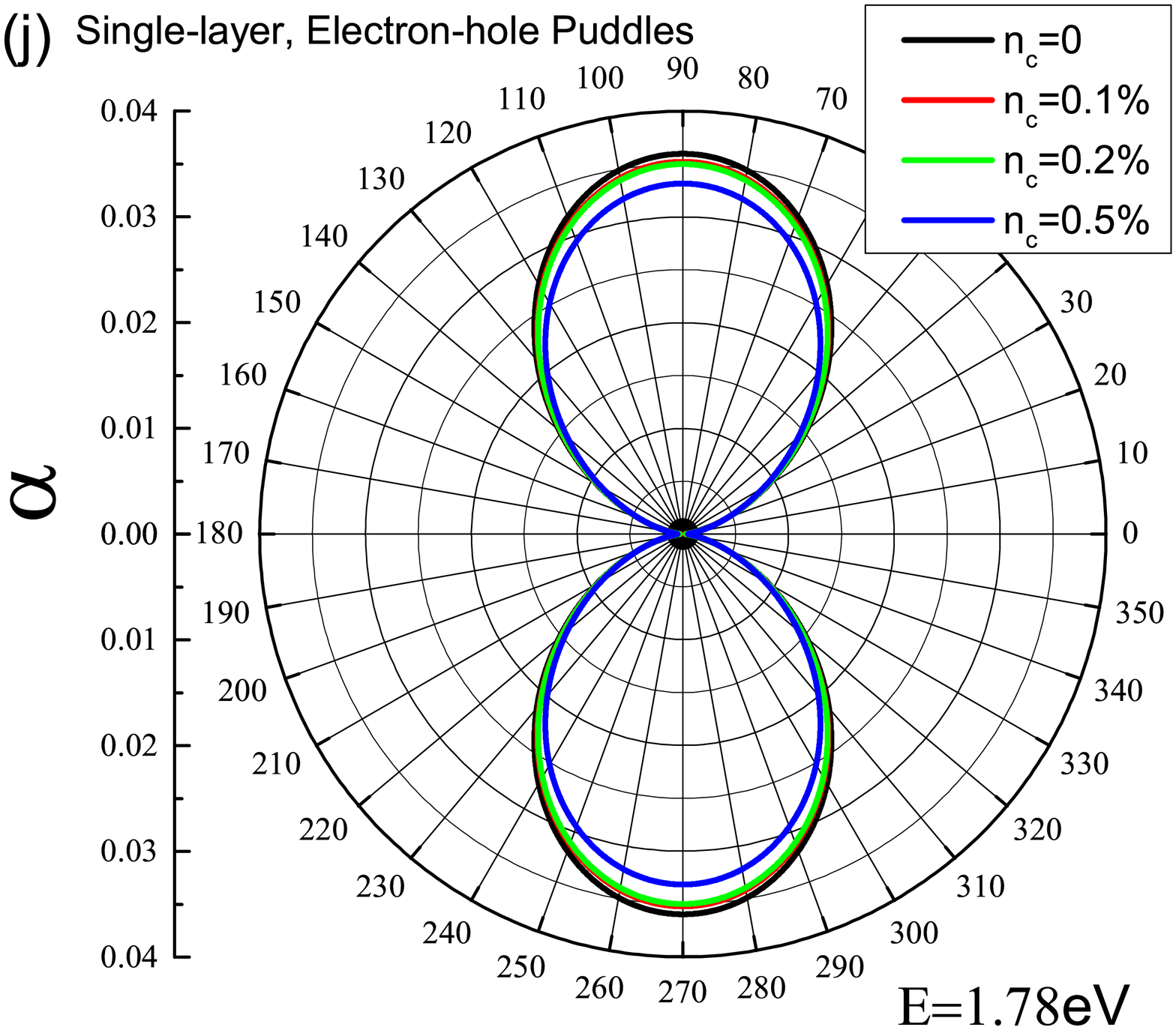}
\includegraphics[width=4.25cm]{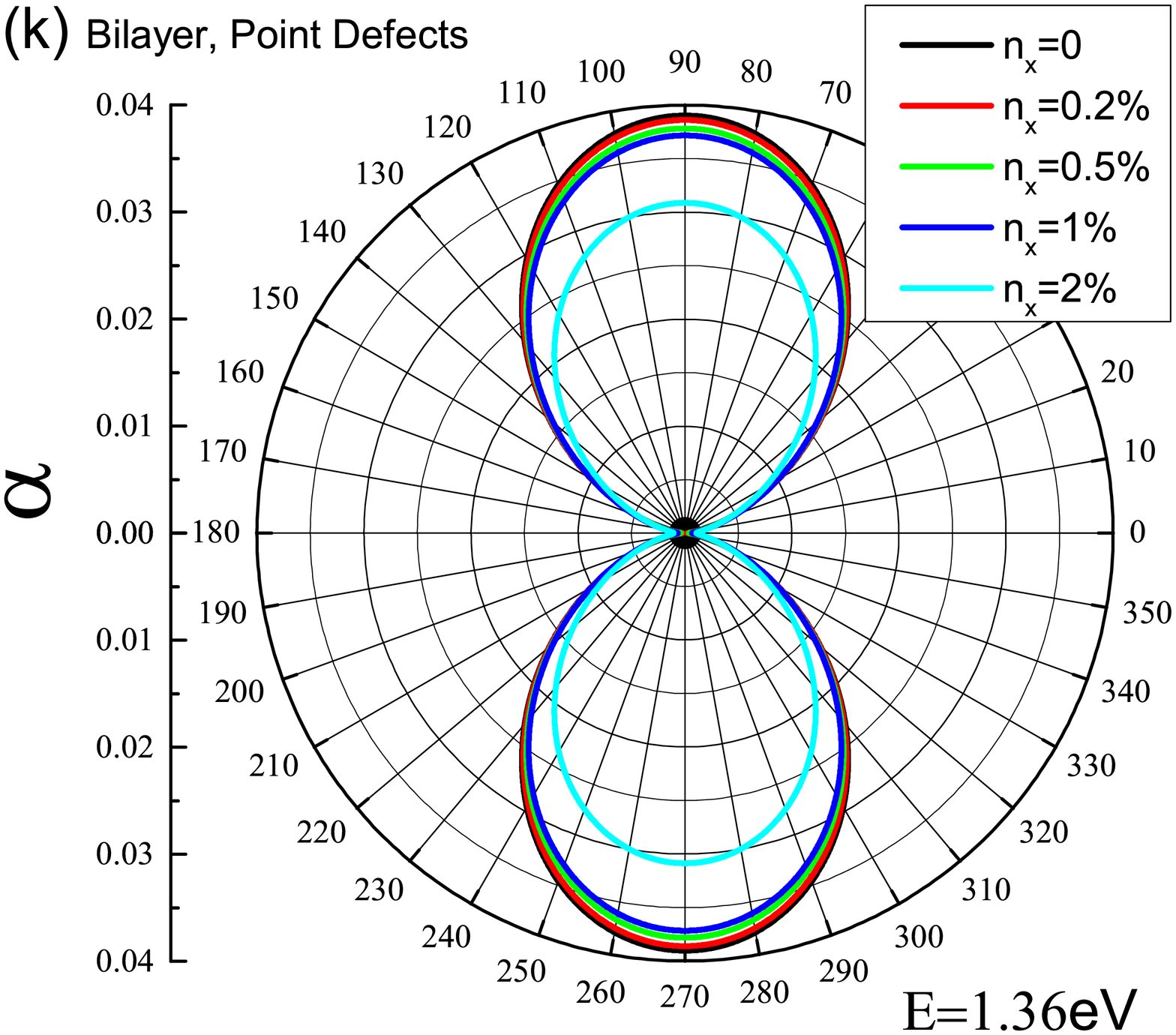}
\includegraphics[width=4.25cm]{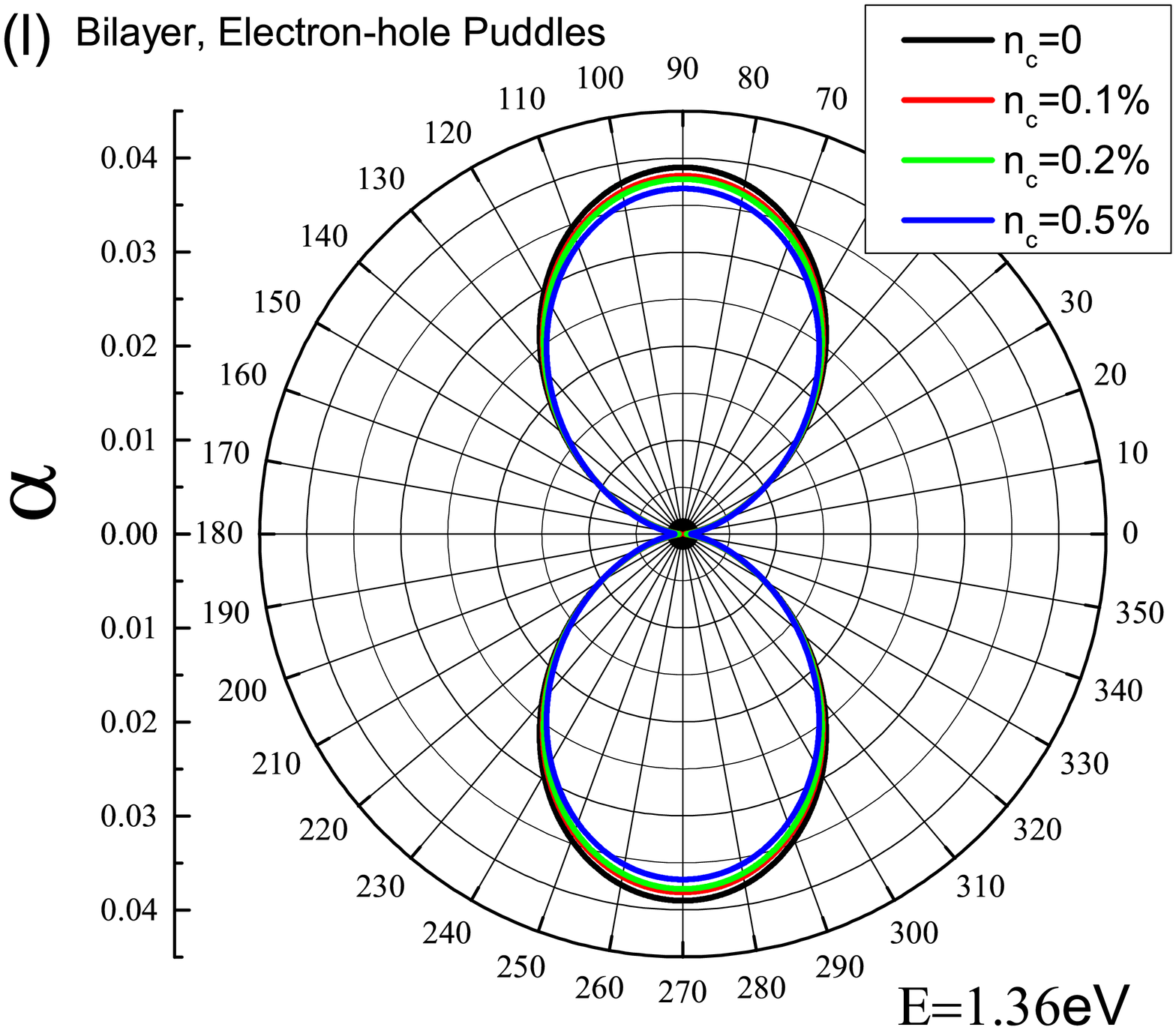}
}
\end{center}
\caption{Optical conductivity (a-d) as a function of energy and optical
absoption (e-l) as a function of polarized angle for suspended single- and
bilayer BP with defects. The energies of the
photon for single-layer are (e,f) $1.4$eV, (i,j) $1.78$eV; and for
bilayer are (j,k) $0.98$eV, (h,l) $1.36$eV. The black lines are the
results for pristine sample, and the color lines are the results for
disordered samples with different concentration of defects.
Throughout this work, we fix the
temperature to $T=300$~K and the chemical potential to $\mu_{F} =0$ for the
optical conductivity, and normalize the conductivity to $\sigma _{0}=\pi
e^{2}/2h$, the universal optical conductivity of single-layer graphene in
the visible light region.}
\label{Fig:AC}
\end{figure*}

Further calculations of the optical conductivity show that there will be
extra excitations below the pristine optical gap along both armchair and
zigzag directions (see Fig. \ref{Fig:AC}), due to the presence of defect
states. Using the presence of point defects as an example, for single-layer,
the excitations between the midgap states at $E\approx -0.2$eV and the
states at the conduction band edge ($E\approx 0.3$eV) reduce the optical gap
to $0.5$eV, much smaller than the value ($1.4$eV) in the pristine sample.
For bilayer, there are two groups of defect states, e.g., one sharp
peak at $E\approx 0.02$eV and another broader peak at $E\approx 0.48$eV
[Fig.~\ref{Fig:DOS}(b)]. The reduced optical gap ($\sim 0.52$eV) is due to
the excitations between the valence band edge ($-0.5$eV) and the conduction
peak at $E\approx 0.02$eV. Another effect due to the appearance of the
defects is the smearing of the optical peaks along the armchair direction,
i.e., the peak at $\omega \approx 1.5$eV for single- and $\omega \approx 1.2$%
eV for bilayer as can be seen from Fig.~\ref{Fig:AC}. It is important to
note that although there are changes of the optical
conductivities along both armchair and zigzag directions, the optical spectrum
has different anisotropy below and above the band gap. The angle-dependent
absorption coefficients of linearly polarized light in Fig. \ref{Fig:AC}
show that the anisotropy remains unchanged for photons with energy higher than
the gap width, but becomes much weaker within the gap. That is, in the
presence of defects, the transport along the armchair direction is
still much stronger than the zigzag direction above the gap, but becomes
comparable within the gap. The difference of the anisotropy is more clear in
the case of single-layer.

The influence of defects on the anisotropy of optical property can be explained
by the isotropic nature of the defect Hamiltonian. A point-like resonant defect is
equivalent to a single lattice site with strong on-site potential or out-of-plane hopping, 
and for a non-resonant defect with real-space Gaussian profile, 
the value of the potential only depends on the
relative distance to the Gaussian center. That is, the extra Hamiltonian terms
introduced by both types of defects are isotropic. Therefore the optical excitations involving the defect states become less
anisotropic compared to excitations in pristine BP. Furthermore, as the new excitation is proportional to the number of defect
states, we expect that 1) the increase of the defect concentration will
enhance the excitation below the optical gap, 2) the profile of the
angle-dependent optical spectroscopy should be robust against the defect
concentration, because the defect states are localized and separated according to the
transport calculations. These conclusions are confirmed by the optical spectroscopy shown in Fig. \ref{Fig:AC}. 

The  restrain of the anisotropy obtained in our calculations is similar as the one observed in
recent excitation measurements of few layer BP films\cite{XiaF2014}. The
difference is that the experiment is performed on a BP film with much
smaller band gap (about $0.3$eV) comparing to these in single-layer ($1.5$%
eV) and bilayer ($1.2$eV). As the band gap in multilayer BP is highly
reduced, the impurity band(s) due to the presence of defects could have
overlaps with the pristine bands, which will restrain the anisotropy even
above the optical gap. We leave out the study of disordered multilayer BP
for future work with a further development of TB models.

\section{Landau level spectrum and magneto-optical spectroscopy}

\begin{figure}[b]
\begin{center}
\mbox{
\includegraphics[width=6cm]{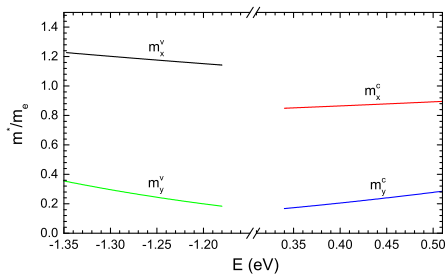}
}
\end{center}
\caption{The energy dependence of the anisotropic effective masses of
single-layer BP in the TB model.}
\label{Fig:Mass}
\end{figure}

\begin{figure}[t]
\begin{center}
\mbox{
\includegraphics[width=8cm]{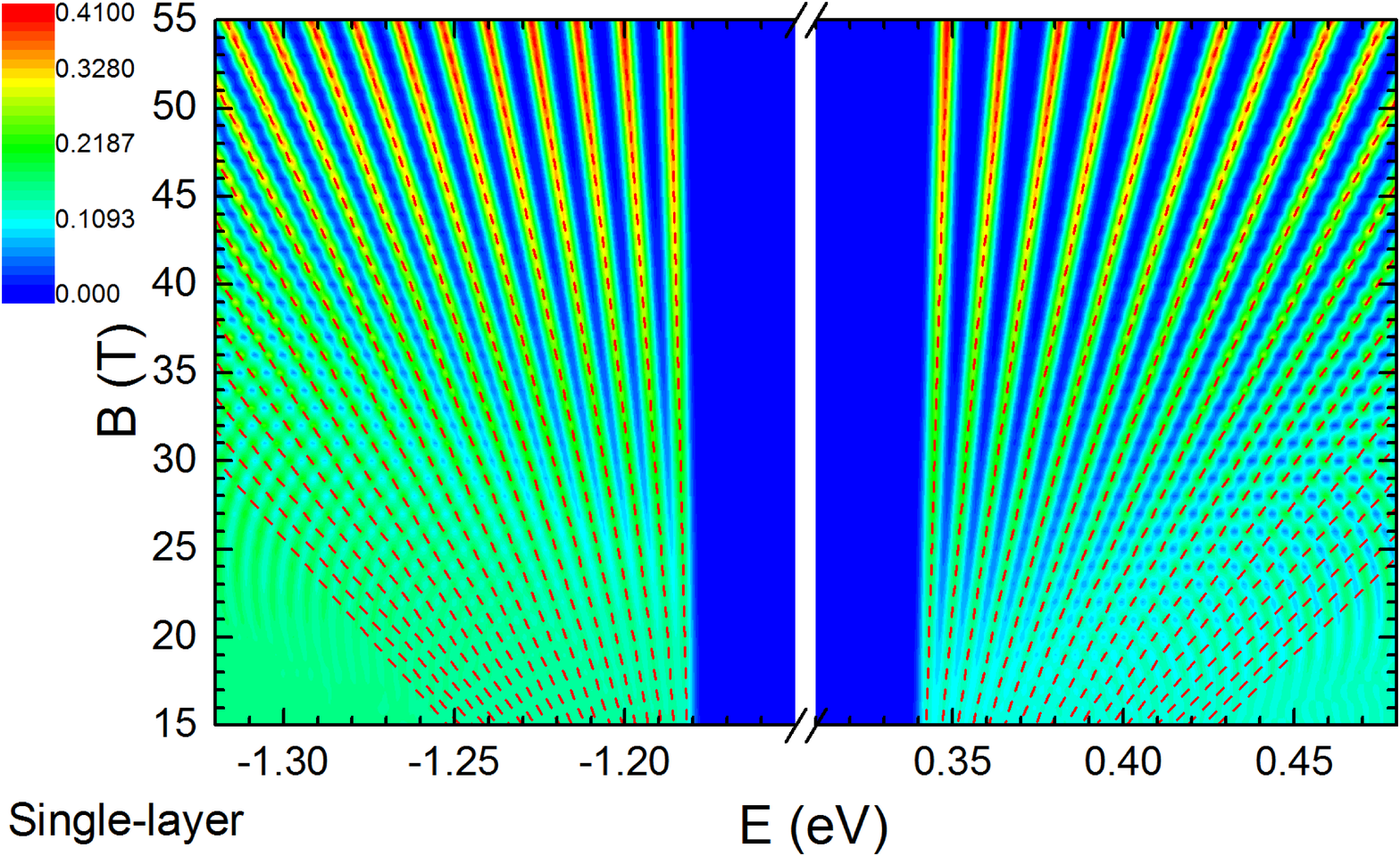}
} 
\mbox{
\includegraphics[width=8cm]{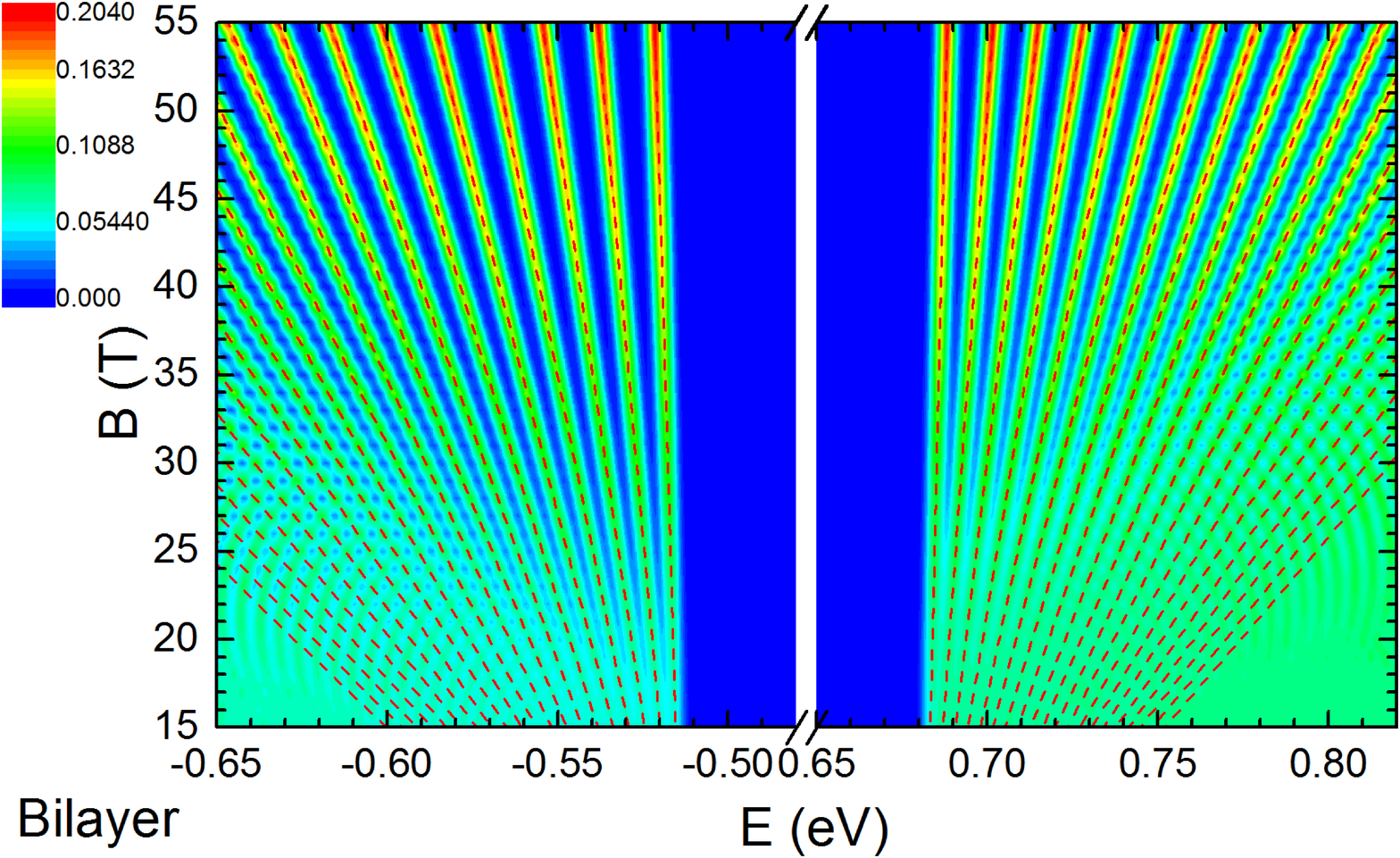}
} 
\end{center}
\caption{Landau level spectrum of single-layer and bilayer BP
in high magnetic field. The red dashed lines are the lowest twenty LLs calculated from Eq. (\protect\ref{Eq:LL_TB}).}
\label{Fig:LandauSpectrum}
\end{figure}

\begin{figure*}[t]
\begin{center}
\mbox{
\includegraphics[width=4.25cm]{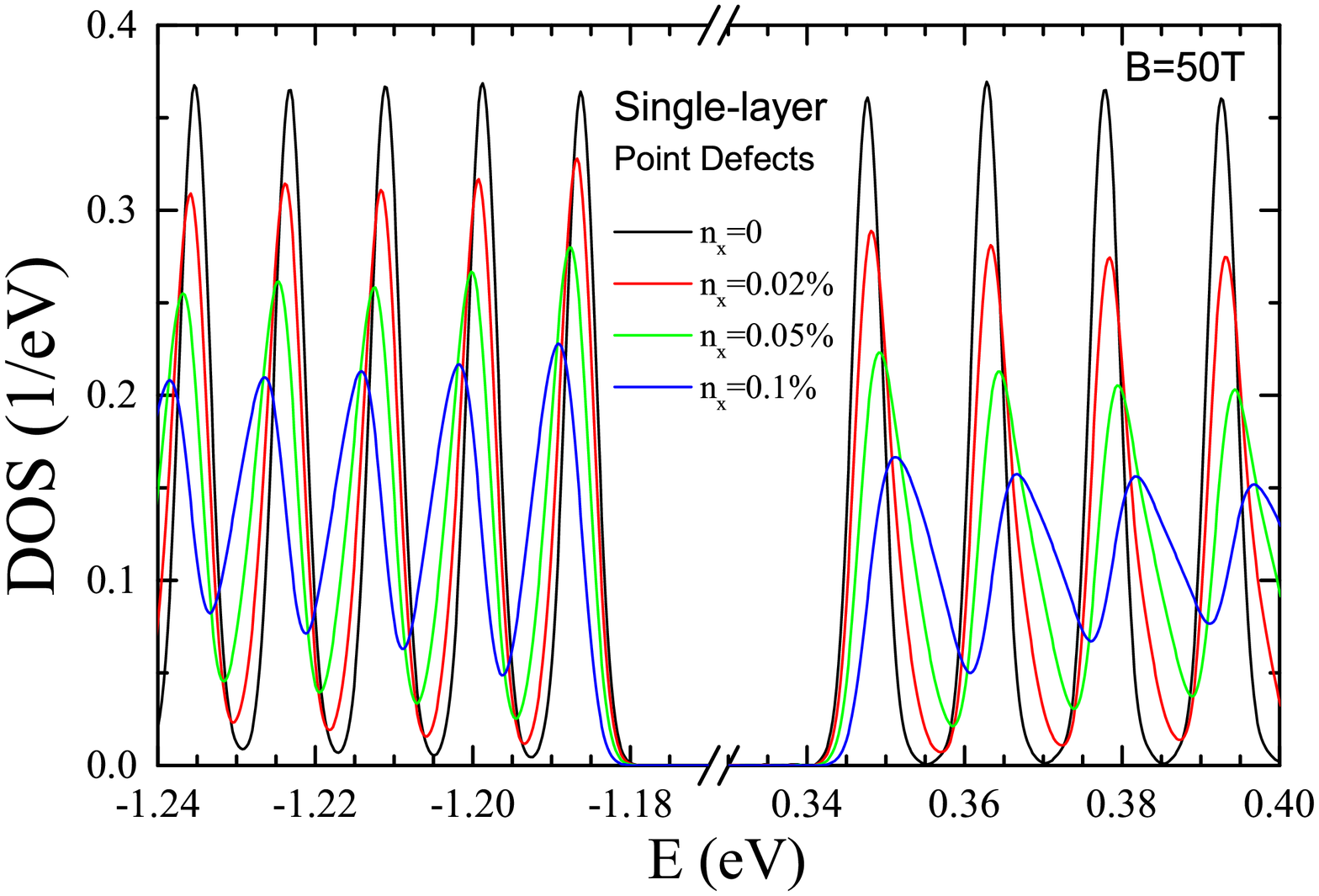}
\includegraphics[width=4.25cm]{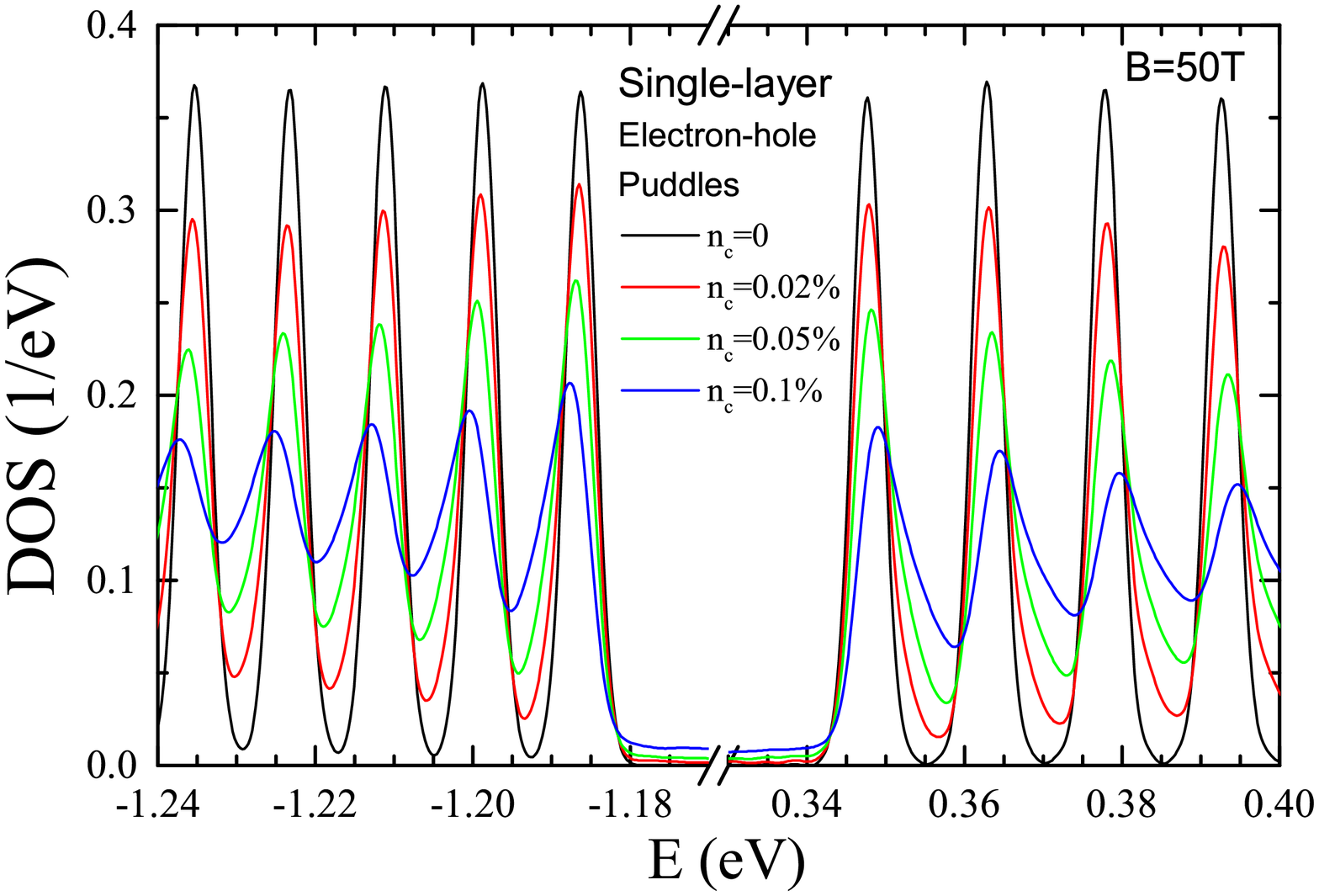}
\includegraphics[width=4.25cm]{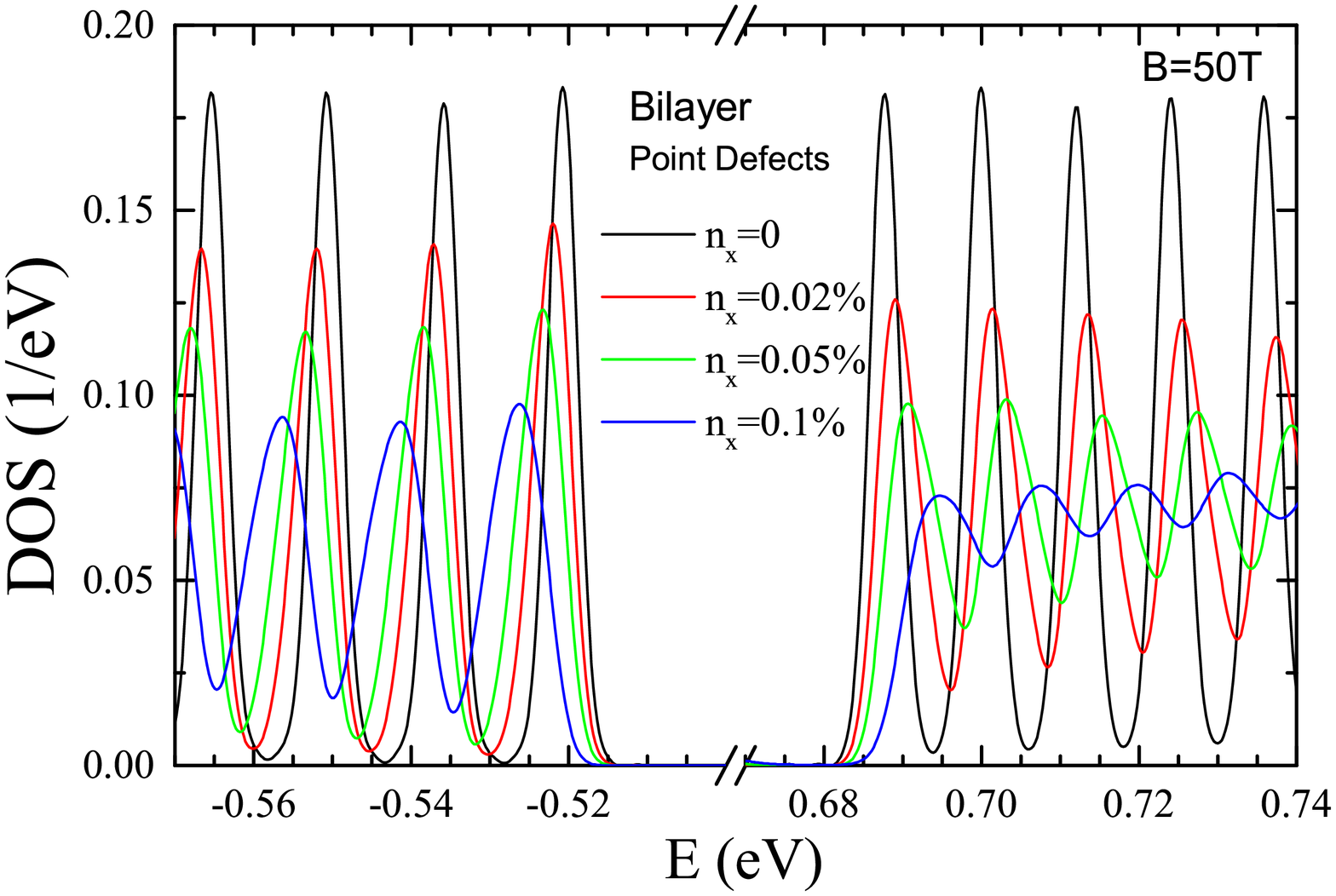}
\includegraphics[width=4.25cm]{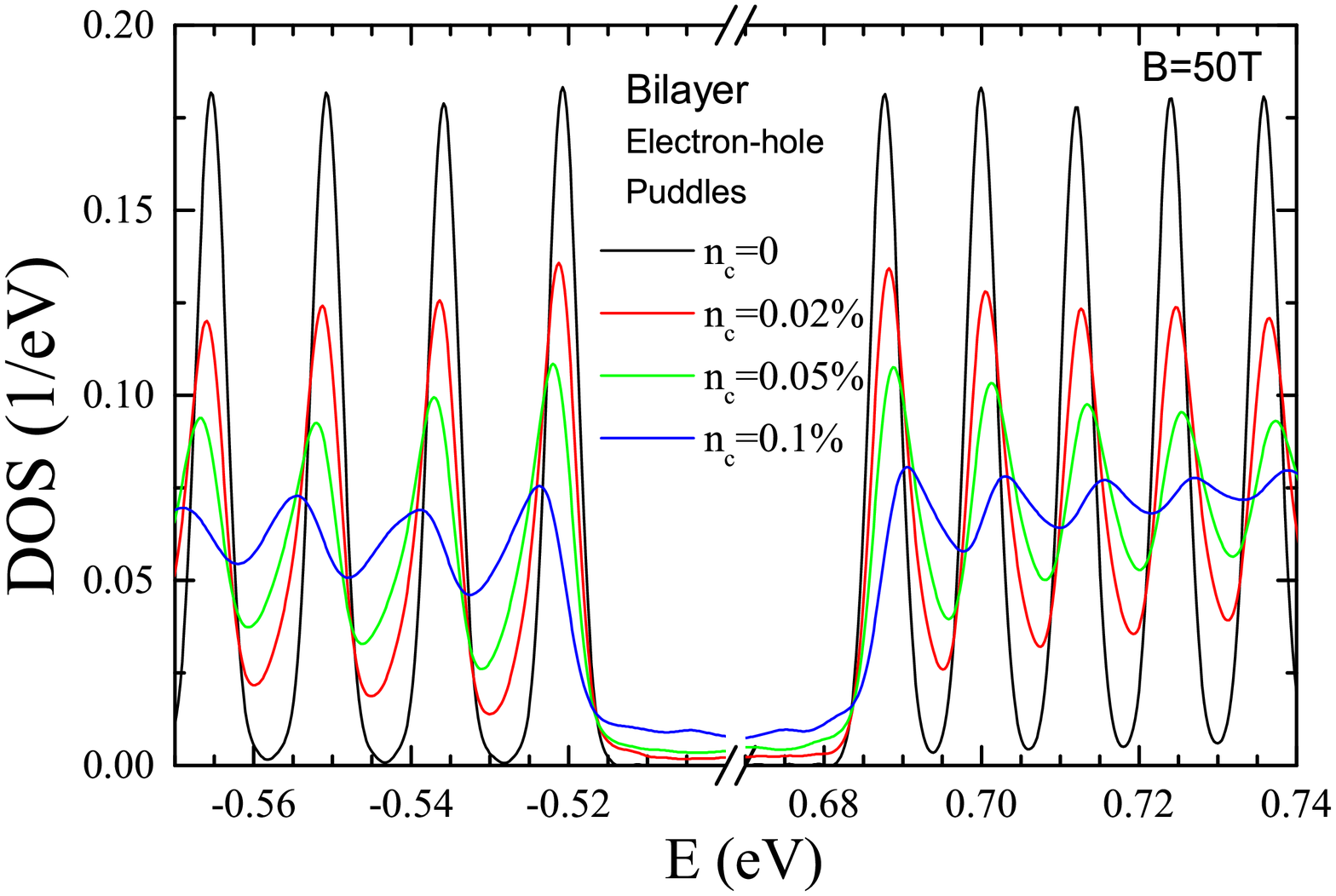}
} 
\mbox{
\includegraphics[width=4.25cm]{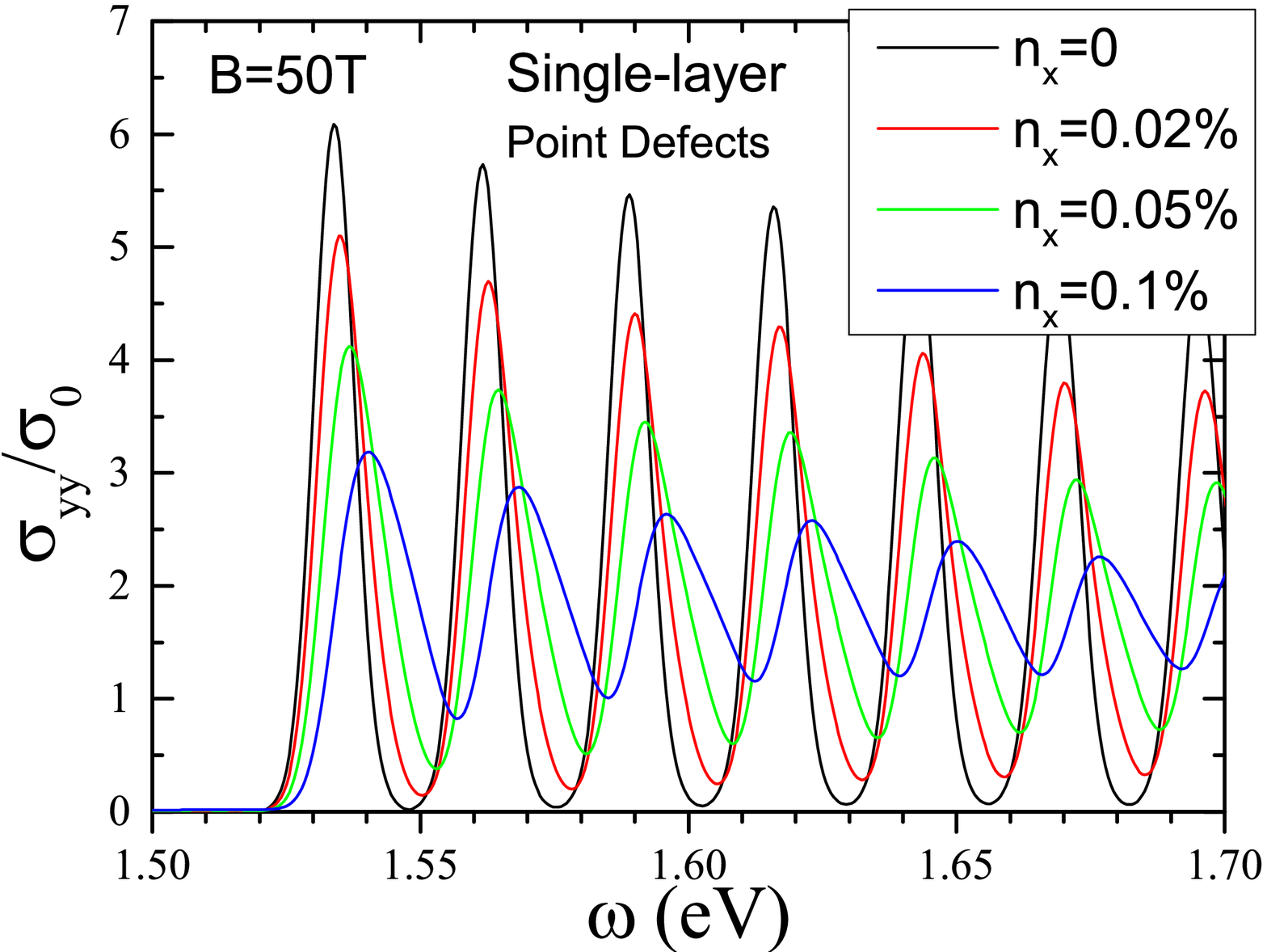}
\includegraphics[width=4.25cm]{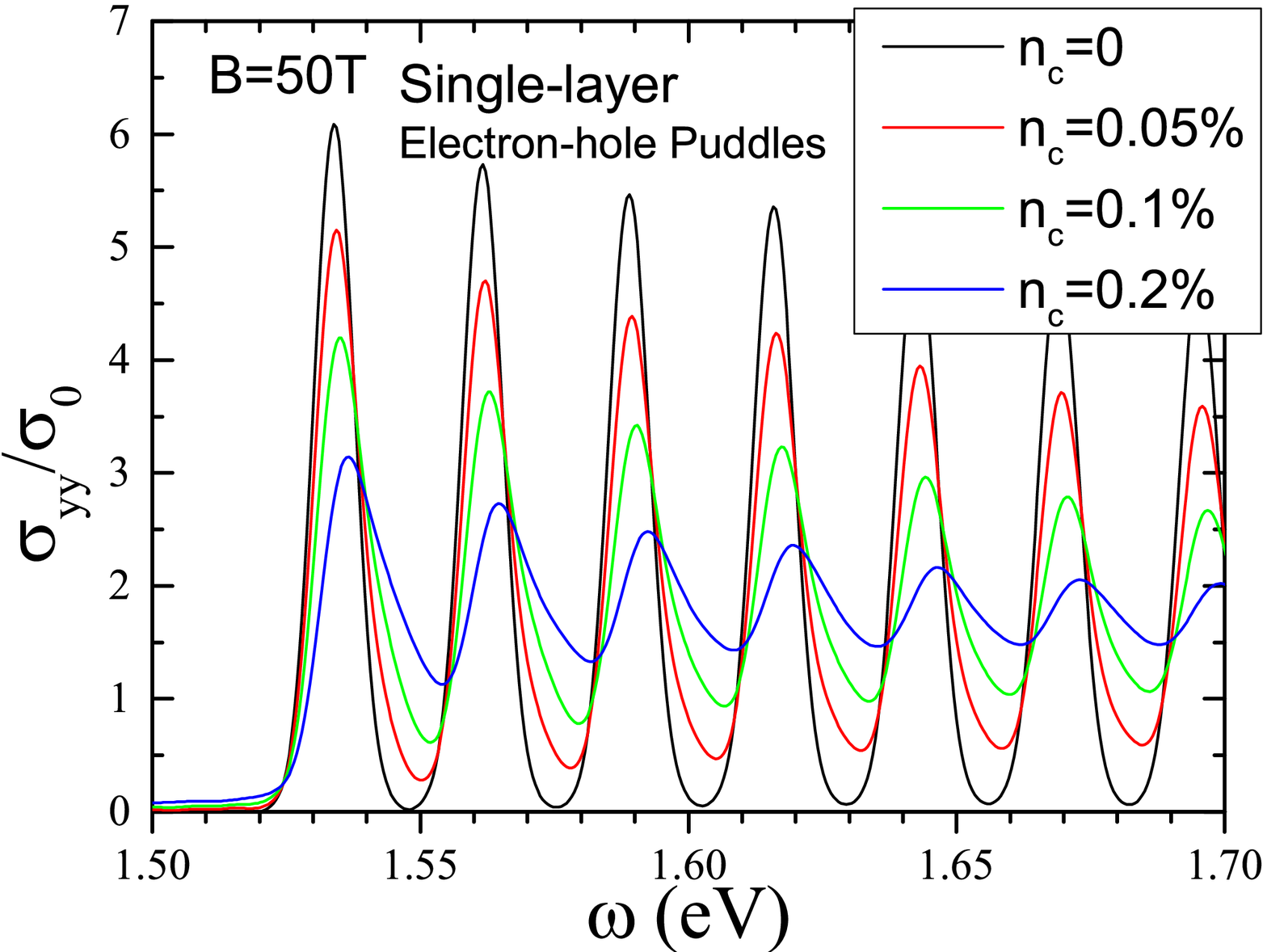}
\includegraphics[width=4.25cm]{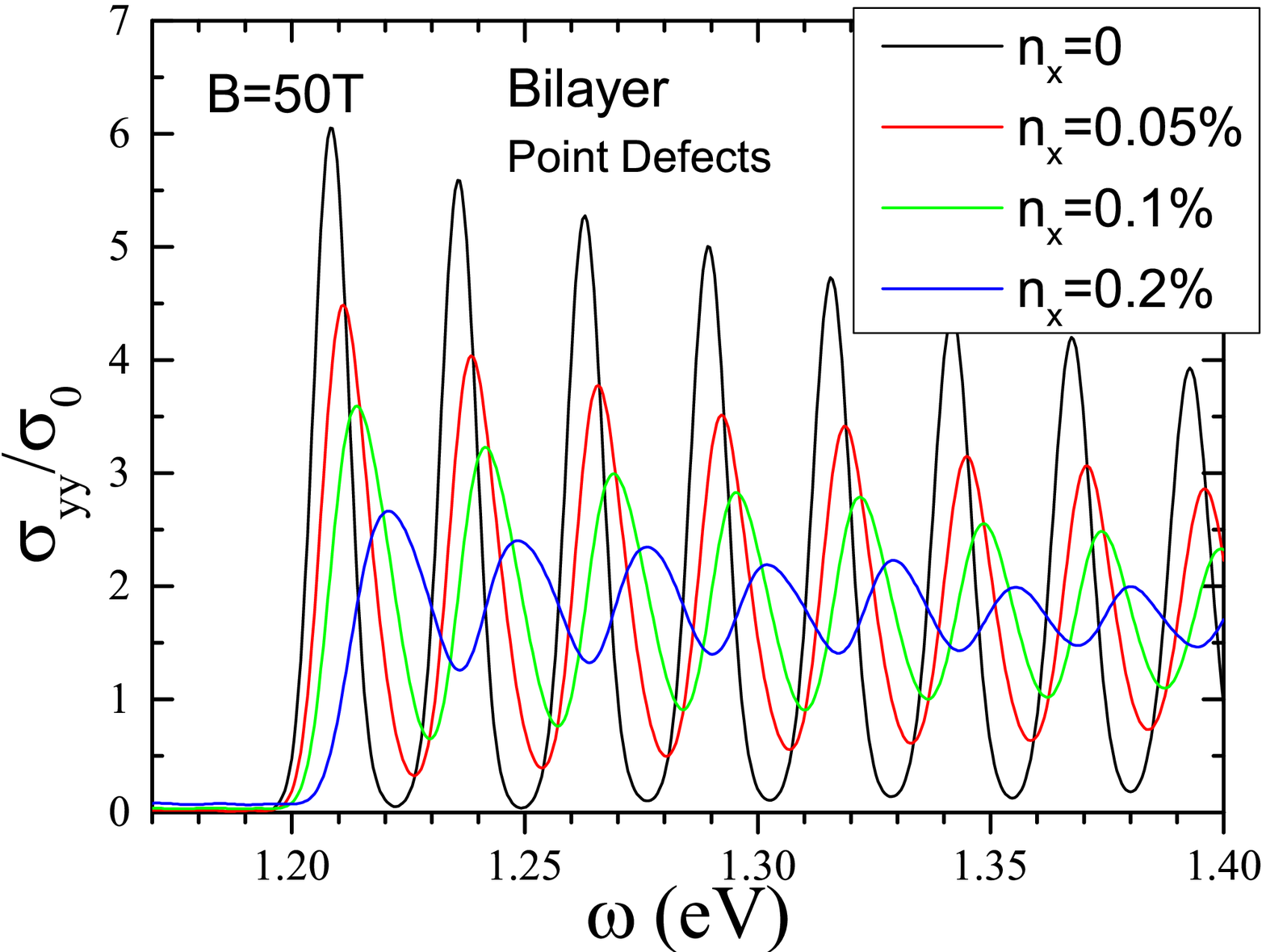}
\includegraphics[width=4.25cm]{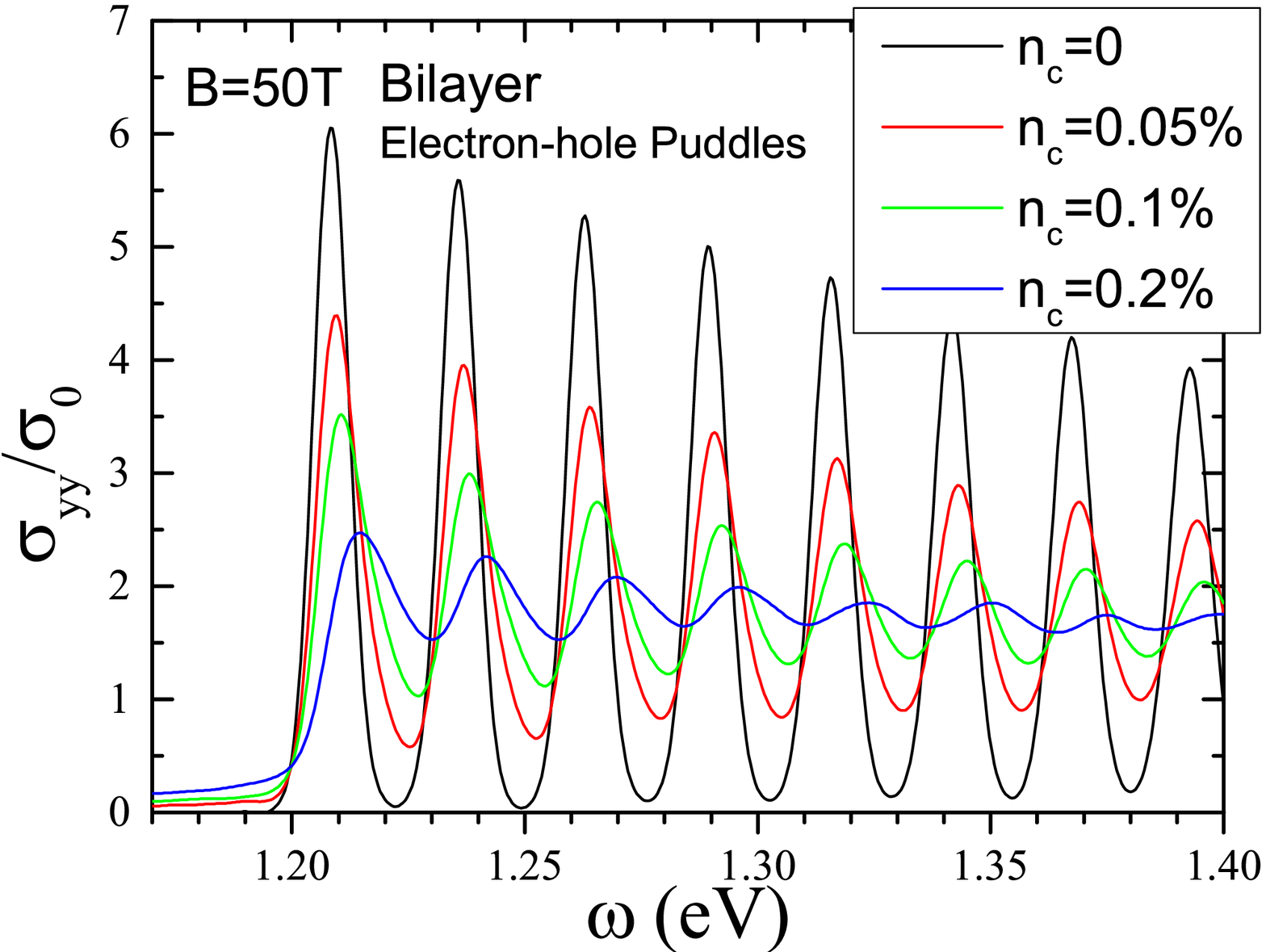}
}
\end{center}
\caption{DOS and magneto-optical spectrum of disordered single-layer and bilayer BP
with perpendicular magnetic field $B=50T$.}
\label{Fig:MagnetoOptical}
\end{figure*}

In the presence of a perpendicular magnetic field $B$, the quantization of the energy levels
leads to separated Landau levels (LLs).
The low energy physics of single-layer BP can be described by
an effective $k\cdot p$ model\cite{Rodin2014,Ezawa2014}, and the LLs follow \cite%
{ZhouXY2014}:%
\begin{equation}
E_{n,s}^{kp}=E_{s}+\frac{seB\hbar }{m_{e}}\left( n+\frac{1}{2}\right) w_{s},
\label{Eq:LLs}
\end{equation}%
where $s=\pm 1$ denotes the conduction and valence bands, $E_{+/-}=E_{c/v}$
the energy at the conduction and valence edge, $n$ the energy index{\ and $%
w_{+/-}=m_{e}/(m_{x}^{c/v}m_{y}^{c/v})^{1/2}$ ($m_{x}^{c/v}$ and $m_{y}^{c/v}
$ are anisotropic effective masses at the $\Gamma $ point). However, as the $%
k\cdot p$ model does not capture the energy-dependence of the effective masses (see 
the results obtained from the TB calculations in Fig. \ref{Fig:Mass}), Eq. (%
\ref{Eq:LLs}) is not valid at high magnetic field. In fact, as can be seen from Fig. \ref{Fig:LandauSpectrum},
the Landau spectrum obtained from the TB calculations 
show that the LLs follow a sublinear dependence on the magnetic field and energy index n,
which can be fitted as
\begin{equation}
E_{n,s}=E_{s}+\frac{seB\hbar }{m_{e}}\left[ \left( n+\frac{1}{2}\right)
w_{s}-\delta _{s}Bn^{p_{s}}\right] ,  \label{Eq:LL_TB}
\end{equation}%
where $E_{+(-)}=0.34eV$ ($-1.18eV$), $w_{+(-)}=2.656$ ($2.181$), $\delta
_{+(-)}=0.0005$ ($0.0004$) and $p_{+/-}=1.8$ for single-layer and $%
E_{+(-)}=0.6815eV$ ($-0.513eV$), $w_{+(-)}=2.14$ ($2.67$), $\delta
_{+(-)}=0.0004$ ($0.00085$) and $p_{+(-)}=1.8$ ($1.73$) for bilayer. 
In the TB calculations, the hopping parameter $t_{mn}$ between
two sites is replaced by a Peierls substitution as $t_{mn}\exp \left[
ie\int_{m}^{n}\mathbf{A}\cdot d\mathbf{l}\right] $, and we choose the Landau
gauge that the vector potential follows $\mathbf{A}=(-By,0,0)$.
In the presence of either point defects or electron-hole
puddles, the LL peaks in the DOS are smeared and suppressed, depending on the
defect concentration. The broadening of peaks in
the DOS also lead to energy shifts of the LLs, however, for small concentration 
of defects, the shifted LLs still
follow the sublinear dependence of Eq. (\ref{Eq:LL_TB}), but with a smaller $\delta _{s}$.
The Landau quantization of energy levels are also observed in the discrete 
magneto-optical spectrum shown in Fig. \ref{Fig:LandauSpectrum}.
The broadening and energy shifts of the optical peaks due to the presence of defects are consistent
with the DOS.
Here we only present the results along the armchair direction, as the magneto-optical conductivity
along the zigzag direction is three-order smaller due to the anisotropy of BP.

\section{Conclusion and Discussion}

In summary, we show that the intrinsic anisotropy of single- and bilayer BP
is robust to the presence of defects. The emergence of defect states with
short-range point defects is identified by sharp peaks in the DOS within the band gap,
which is different from the uniform increase of states with the long-range electron-hole puddles. For
both short- and long-range defects, the defect states are
insulating due to the Anderson localization in disordered 2D system,
but they cause extra excitations within the optical gap. The dc conductivity
as well as carrier mobility beyond the gap are significantly reduced due to
the scattering from the defects. The angle-dependent absorption coefficients
of linearly polarized light show that the anisotropy above the band gap are
robust against the disorder, but the anisotropy of the new excitations
involving the defect states are suppressed, because of the isotropic nature
of the defects. By fitting the
numerical results of the DOS obtained in the TB model, we find a sublinear
dependence of LLs on the magnetic field and level index, even at low defect concentrations.

\section{Acknowledgments}

We thank J. M. Pereira Jr. for useful discussions. The support by the
Stichting Fundamenteel Onderzoek der Materie (FOM) and the Netherlands
National Computing Facilities foundation (NCF) are acknowledged. We thank
financial support from the European Research Council Advanced Grant program
(contract 338957). The research has also received funding from the European Union
Seventh Framework Programme under Grant Agreement No. 604391 Graphene
Flagship.

\begin{widetext}
\section{Appendix}

The unit cell of single-layer BP contains four atoms, and the TB Hamiltonian
can be represented as \cite{Rudenko2014,Ezawa2014}%
\begin{eqnarray}
H &=&\left( 
\begin{array}{cccc}
0 & t_{2}\varphi _{2}+t_{5}\varphi _{5} & t_{4}\varphi _{4} & t_{1}\varphi
_{1}+t_{3}\varphi _{3} \\ 
t_{2}\varphi _{2}^{\ast }+t_{5}\varphi _{5}^{\ast } & 0 & t_{1}\varphi
_{1}^{\ast }+t_{3}\varphi _{3}^{\ast } & t_{4}\varphi _{4}^{\ast } \\ 
t_{4}\varphi _{4}^{\ast } & t_{1}\varphi _{1}+t_{3}\varphi _{3} & 0 & 
t_{2}\varphi _{2}+t_{5}\varphi _{5} \\ 
t_{1}\varphi _{1}^{\ast }+t_{3}\varphi _{3}^{\ast } & t_{4}\varphi _{4} & 
t_{2}\varphi _{2}^{\ast }+t_{5}\varphi _{5}^{\ast } & 0%
\end{array}%
\right)  \notag  \label{HamiltonianMatrix} \\
&&
\end{eqnarray}%
where the phase terms $\varphi _{i}$ are defined as%
\begin{eqnarray}
\varphi _{1} &=&2e^{idk_{y}}\cos \left( ck_{x}\right) , \\
\varphi _{2} &=&e^{-ibk_{y}}, \\
\varphi _{3} &=&2e^{-i(2b+d)k_{y}}\cos \left( ck_{x}\right) , \\
\varphi _{4} &=&4\cos \left( ck_{x}\right) \cos \left( (b+d)k_{y}\right) , \\
\varphi _{5} &=&e^{i(b+2d)k_{y}},
\end{eqnarray}%
and the constants $c=a\sin \theta $, $d=a\cos \theta $, $\theta
=48.395^{0},\ a\approx 2.216$\AA\ and $b\approx 0.716$\AA\ are the atomic
distance of two nearest neighbors projected to the surface plane. The four
eigenvalues of the TB Hamiltonian matrix can be represented as 
\begin{eqnarray}
E_{1}(k_{x},k_{y}) &=&A(k_{x},k_{y})-B(k_{x},k_{y}), \\
E_{2}(k_{x},k_{y}) &=&A(k_{x},k_{y})+B(k_{x},k_{y}), \\
E_{3}(k_{x},k_{y}) &=&-A(k_{x},k_{y})-C(k_{x},k_{y}), \\
E_{4}(k_{x},k_{y}) &=&-A(k_{x},k_{y})+C(k_{x},k_{y}),
\end{eqnarray}%
where $A(k_{x},k_{y})$, $B(k_{x},k_{y})$ and $C(k_{x},k_{y})$ are%
\begin{eqnarray}
A(k_{x},k_{y}) &=&4t_{4}\cos \left( ck_{x}\right) \cos \left(
(b+d)k_{y}\right) ,  \notag \\
B(k_{x},k_{y}) &=&[2\cos \left( 2ck_{x}\right) \left( 2t_{1}t_{3}\cos \left(
2(b+d)k_{y}\right) +t_{1}^{2}+t_{3}^{2}\right) +4\cos \left( ck_{x}\right)
((t_{2}(t_{1}+t_{3})+t_{1}t_{5})\cos \left( (b+d)k_{y}\right)  \notag \\
&&+t_{3}t_{5}\cos \left( 3(b+d)k_{y}\right) +2(2t_{1}t_{3}+t_{2}t_{5})\cos
\left( 2(b+d)k_{y}\right) +2t_{1}^{2}+t_{2}^{2}+2t_{3}^{2}+t_{5}^{2}]^{1/2}.
\notag \\
C(k_{x},k_{y}) &=&[-4t_{3}t_{5}\cos \left( ck_{x}\right) \cos \left(
3(b+d)k_{y}\right) +2\cos \left( 2(b+d)k_{y}\right) \left( 2t_{1}t_{3}\cos
\left( 2ck_{x}\right) +2t_{1}t_{3}+t_{2}t_{5}\right)  \notag \\
&&-4\left( t_{2}t_{3}+t_{1}\left( t_{2}+t_{5}\right) \right) \cos \left(
ck_{x}\right) \cos \left( (b+d)k_{y}\right) +2\left(
t_{1}^{2}+t_{3}^{2}\right) \cos \left( 2ck_{x}\right)
+2t_{1}^{2}+t_{2}^{2}+2t_{3}^{2}+t_{5}^{2}]^{1/2}  \notag \\
&&
\end{eqnarray}%
$E_{1}(k_{x},k_{y})$ and $E_{2}(k_{x},k_{y})$ are the lowest valence ($E^{v}$%
) and conduction ($E^{c}$) bands fitted to GW calculations, and plotted in
Fig.~\ref{Fig:Hoppings_Band}.

The Fermi velocity can be obtained via $v_{\alpha }=\frac{1}{\hbar }\frac{%
\partial E}{\partial k_{\alpha }},$ and the electron and hole velocities
along the armchair (Y) and zigzag (X) directions are%
\begin{eqnarray}
v_{x}^{v} &=&A_{1}-B_{1}/D, \\
v_{y}^{v} &=&A_{2}-B_{2}/D, \\
v_{x}^{c} &=&A_{1}+B_{1}/D, \\
v_{y}^{c} &=&A_{2}+B_{2}/D,
\end{eqnarray}%
where%
\begin{eqnarray*}
A_{1} &=&-4ct_{4}\sin \left( ck_{x}\right) \cos \left( (b+d)k_{y}\right) , \\
A_{2} &=&-4t_{4}(b+d)\cos \left( ck_{x}\right) \sin \left( (b+d)k_{y}\right)
, \\
B_{1} &=&-4c\sin \left( 2ck_{x}\right) \left( 2t_{3}t_{1}\cos \left(
2(b+d)k_{y}\right) +t_{1}^{2}+t_{3}^{2}\right) \\
&&-4c\sin \left( ck_{x}\right) \left( t_{3}t_{5}\cos \left(
3(b+d)k_{y}\right) +\left( t_{2}\left( t_{1}+t_{3}\right) +t_{1}t_{5}\right)
\cos \left( (b+d)k_{y}\right) \right) , \\
B_{2} &=&-8t_{1}t_{3}(b+d)\cos \left( 2ck_{x}\right) \sin \left(
2(b+d)k_{y}\right) -4\left( 2t_{1}t_{3}+t_{2}t_{5}\right) (b+d)\sin \left(
2(b+d)k_{y}\right) \\
&&+4\cos \left( ck_{x}\right) \left( -3t_{3}t_{5}(b+d)\sin \left(
3(b+d)k_{y}\right) -\left( t_{2}\left( t_{1}+t_{3}\right) +t_{1}t_{5}\right)
(b+d)\sin \left( (b+d)k_{y}\right) \right) \\
D &=&2[2\cos \left( 2ck_{x}\right) \left( 2t_{3}t_{1}\cos \left(
2(b+d)k_{y}\right) +t_{1}^{2}+t_{3}^{2}\right) \\
&&+4\cos \left( ck_{x}\right) \left( t_{3}t_{5}\cos \left(
3(b+d)k_{y}\right) +\left( t_{2}\left( t_{1}+t_{3}\right) +t_{1}t_{5}\right)
\cos \left( (b+d)k_{y}\right) \right) \\
&&+2\left( 2t_{1}t_{3}+t_{2}t_{5}\right) \cos \left( 2(b+d)k_{y}\right)
+2t_{1}^{2}+t_{2}^{2}+2t_{3}^{2}+t_{5}^{2}]^{1/2}.
\end{eqnarray*}

The calculation of effective mass is straight forward via $1/m_{\alpha }=%
\frac{\partial ^{2}E }{\partial k_{\alpha }^{2}}/\hbar ^{2}$, and for $%
k_{x}=k_{y}=0$\thinspace , we have%
\begin{eqnarray}
m_{x}^{v} &=&-4c^{2}t_{4}-A_{3}/F, \\
m_{y}^{v} &=&-4t_{4}(b+d)^{2}-A_{4}/F, \\
m_{x}^{c} &=&-4c^{2}t_{4}+A_{3}/F, \\
m_{y}^{c} &=&-4t_{4}(b+d)^{2}+A_{4}/F,
\end{eqnarray}%
where%
\begin{eqnarray*}
A_{3} &=&-8c^{2}\left( t_{1}^{2}+2t_{3}t_{1}+t_{3}^{2}\right) -4c^{2}\left(
t_{2}\left( t_{1}+t_{3}\right) +t_{1}t_{5}+t_{3}t_{5}\right) , \\
A_{4} &=&-16t_{1}t_{3}(b+d)^{2}-8\left( 2t_{1}t_{3}+t_{2}t_{5}\right)
(b+d)^{2}+4\left( \left( t_{2}\left( t_{1}+t_{3}\right) +t_{1}t_{5}\right)
(-b-d)(b+d)-9t_{3}t_{5}(b+d)^{2}\right) , \\
F &=&2\sqrt{2t_{1}^{2}+t_{2}^{2}+2t_{3}^{2}+t_{5}^{2}+2\left(
t_{1}^{2}+2t_{3}t_{1}+t_{3}^{2}\right) +2\left(
2t_{1}t_{3}+t_{2}t_{5}\right) +4\left( t_{2}\left( t_{1}+t_{3}\right)
+t_{1}t_{5}+t_{3}t_{5}\right) }.
\end{eqnarray*}%
By using the parameters of single-layer BP, the effective masses at $\Gamma $
point are%
\begin{eqnarray}
m_{x}^{v} &=&-1.143m_{e}, \\
m_{y}^{v} &=&-0.184m_{e}, \\
m_{x}^{c} &=&0.849m_{e}, \\
m_{y}^{c} &=&0.167m_{e}.
\end{eqnarray}

\end{widetext}

\bibliographystyle{apsrev}
\bibliography{Bib_2D}

\end{document}